%% file: TOP-26-001_temp.tex
\documentclass[11pt,twoside,a4paper,cmspaper,final,collab]{cms-tdr}
\def\svnVersion{f7ecb9b-D}\def\svnDate{2026/10/29}\def\cmsCernNoTag{CERN-EP-2026-274}\def\cmsCernDate{\today}\def\cmsMessage{Submitted to the Journal of High Energy Physics}
\begin{document}\cmsNoteHeader{TOP-26-001}

\newcommand{\Wcb}{\ensuremath{\PW\to\PQc\PQb}\xspace}
\newcommand{\Wud}{\ensuremath{\PW\to\PQu\PQd}\xspace}
\newcommand{\Wus}{\ensuremath{\PW\to\PQu\PQs}\xspace}
\newcommand{\Wcd}{\ensuremath{\PW\to\PQc\PQd}\xspace}
\newcommand{\Wcs}{\ensuremath{\PW\to\PQc\PQs}\xspace}
\newcommand{\Wub}{\ensuremath{\PW\to\PQu\PQb}\xspace}
\newcommand{\Wudus}{\ensuremath{\PW\to\PQu\PQd\,(\PQu\PQs)}\xspace}
\newcommand{\Wcscd}{\ensuremath{\PW\to\PQc\PQs\,(\PQc\PQd)}\xspace}
\newcommand{\Dreco}{\ensuremath{\mathcal{D}_{\mathrm{reco}}}\xspace}
\newcommand{\Dcb}{\ensuremath{\mathcal{D}_{\Wcb}}\xspace}
\newcommand{\pp}{\ensuremath{\Pp\Pp}\xspace}
\newcommand{\hdamp}{\ensuremath{h_{\text{damp}}}\xspace}
\newcommand{\VcbAbs}{\ensuremath{\abs{V_{\PQc\PQb}}}\xspace}
\newcommand{\Vcb}{\ensuremath{V_{\PQc\PQb}}\xspace}
\newcommand{\VcbPDG}{\ensuremath{\abs{V_{\PQc\PQb}^{\mathrm{PDG}}}}\xspace}
\newcommand{\tW}{\ensuremath{\PQt{}\PW}\xspace}
\newcommand{\Vjets}{\ensuremath{\PV{+}\text{jets}}\xspace}
\newcommand{\GammaStar}{\ensuremath{\PGg{}^{*}}\xspace}
\newcommand{\ttV}{\ensuremath{\ttbar{}\PV}\xspace}
\newcommand{\ttZ}{\ensuremath{\ttbar{}\PZ}\xspace}
\newcommand{\ttX}{\ensuremath{\ttbar{}\PX}\xspace}
\newcommand{\WW}{\ensuremath{\PW{}\PW}\xspace}
\newcommand{\WZ}{\ensuremath{\PW{}\PZ}\xspace}
\newcommand{\ZZ}{\ensuremath{\PZ{}\PZ}\xspace}
\newcommand{\VV}{\ensuremath{\PV{}\PV}\xspace}
\newcommand{\ttj}{\ensuremath{\ttbar{+}\text{light}}\xspace}
\newcommand{\tthf}{\ensuremath{\ttbar{+}\text{HF}}\xspace}
\newcommand{\ttbb}{\ensuremath{\ttbar\bbbar}\xspace}
\newcommand{\POWHEGBOX}{{\POWHEG}\textsc{-box-res}\xspace}
\newcommand{\OPENLOOPS}{\textsc{OpenLoops}\xspace}
\newcommand{\jhadtb}{\ensuremath{\vphantom{j_{\PQb}}\smash{j_{\PQb}^{\PQt_{\smash[b]{\text{had}}}}}}\xspace}
\newcommand{\jwu}{\ensuremath{j_{\mathrm{U}}^{\PW}}\xspace}
\newcommand{\jwd}{\ensuremath{j_{\mathrm{D}}^{\PW}}\xspace}
\newcommand{\jleptb}{\ensuremath{\vphantom{j_{\PQb}}\smash{j_{\PQb}^{\PQt_{\smash[b]{\text{lep}}}}}}\xspace}
\newcommand{\pzN}{\ensuremath{p_{z}^{\PGn}}\xspace}
\newcommand{\bvsc}{\ensuremath{B}\xspace}
\newcommand{\minRbb}{\ensuremath{\min \Delta R_{\PQb\PQb}}\xspace}
\newcommand{\minMbb}{\ensuremath{\min m_{\PQb\PQb}}\xspace}
\newcommand{\muR}{\ensuremath{\mu_{\mathrm{R}}}\xspace}
\newcommand{\muF}{\ensuremath{\mu_{\mathrm{F}}}\xspace}
\newlength\cmsTabSkip
\setlength\cmsTabSkip{1ex}
\cmsNoteHeader{TOP-26-001}
\title{Measurement of the \texorpdfstring{\VcbAbs}{|Vcb|} element of the CKM matrix using \texorpdfstring{\ttbar}{ttbar} events in \texorpdfstring{\pp}{pp} collisions at \texorpdfstring{$\sqrt{s}=13\TeV$}{sqrt(s) = 13 TeV}}

\date{\today}

\abstract{A measurement of the magnitude of the Cabibbo--Kobayashi--Maskawa matrix element \Vcb is presented using top quark pair events produced in proton-proton collisions at $\sqrt{s}=13\TeV$. The data correspond to an integrated luminosity of 138\fbinv collected by the CMS experiment at the LHC. The measurement targets the lepton$+$jets final state, where one \PW boson decays leptonically and the other hadronically. The value of \VcbAbs is extracted from the branching fraction of the \Wcb decay. To identify this rare signal amid the dominant \Wud and \Wcs backgrounds, a multivariate discriminant based primarily on the \PQb jet tagging information of the jets assigned to the hadronic \PW boson decay is used. The measured value is $\VcbAbs=[4.59^{+0.44}_{-0.48}\stat^{+0.64}_{-0.76}\syst]\times 10^{-2}$. This measurement provides a complementary determination of \VcbAbs at the electroweak scale, independent of the nonperturbative quantum chromodynamics inputs used in measurements based on semileptonic \PB meson decays.}

\hypersetup{
  pdfauthor={CMS Collaboration},
  pdftitle={Measurement of the |Vcb| element of the CKM matrix using ttbar events in pp collisions at  sqrt(s) = 13 TeV},
  pdfsubject={CMS},
  pdfkeywords={CKM matrix, Vcb, top quark}}

\maketitle

\section{Introduction}\label{sec:introduction}
The flavor-changing charged current interactions of quarks are governed by the Cabibbo--Kobayashi--Maskawa (CKM) matrix.
Precise determination of the CKM matrix elements is one of the important tests of the standard model (SM).
Inconsistencies among the measurements could indicate the presence of physics beyond the SM.

Currently, the determination of \VcbAbs relies primarily on measurements of semileptonic \PB meson decays.
These measurements are classified into two distinct approaches.
The first method combines measurements of the inclusive decay rate and moments of the kinematic distributions in $\PB \to X_{\PQc} \Pell \PAGnell$, where $X_{\PQc}$ denotes any charmed hadronic final state.
The extraction of \VcbAbs then relies on the heavy-quark expansion~\cite{BORDONE2021136679, Bernlochner2022}.
In contrast, the second method is based on exclusive decays, such as $\PB \to \PD^{(*)} \Pell \PAGnell$, which depend on hadronic form factors parameterized using experimental data and constrained by lattice quantum chromodynamics (QCD) calculations~\cite{PhysRevLett.123.091801, PhysRevD.108.092013, PhysRevLett.133.131801, PhysRevLett.104.011802, PhysRevD.93.032006, vs8k-259v}.
Despite significant experimental and theoretical advances, a long-standing discrepancy exceeding three standard deviations persists between the values obtained from these two methods~\cite{ParticleDataGroup:2026}.
This unresolved tension is widely referred to as the \VcbAbs puzzle.

To resolve this discrepancy, an independent determination of \VcbAbs is highly desirable.
The idea of extracting \VcbAbs from the hadronic $\PWp\to\PQc\PAQb$ and $\PWm\to\PAQc\PQb$ decays in top quark pair (\ttbar) events at the CERN LHC was initially proposed in Ref.~\cite{Harrison2019}.
Compared with semileptonic \PB meson decays, which probe \VcbAbs at a characteristic scale of order $m_{\PQb}/2$, this approach probes it near the electroweak scale, of order $m_{\PW}$.
This approach does not rely on the modeling of low-$q^2$ hadronic form factors, heavy-quark expansion inputs, or meson bound-state dynamics.
It is therefore subject to distinct theoretical and experimental inputs, making it complementary to \PB meson decay measurements.
In what follows, charge-conjugate states are implied, and the notation $\PW\to\PQq\PQq'$ is used as shorthand for the corresponding pair of decay modes.

The original proposal in Ref.~\cite{Harrison2019} estimated the sensitivity using event counts and simplified assumptions about event selection and flavor tagging.
A subsequent study in Ref.~\cite{Choi2021} incorporated fast detector simulation and kinematic reconstruction of \ttbar events, but used only the number of \PQb-tagged jets assigned to the hadronic \PW boson decay as its discriminant, leaving room for improved sensitivity.
Using a distinct muon-in-jet charm-tagging technique, a related CMS measurement determined the inclusive fraction of hadronic \PW boson decays containing a \PQc quark in \ttbar events to be $0.489\pm0.020$~\cite{CMS:SMP-24-009}.
A measurement of \VcbAbs in \ttbar decays has recently been reported by the ATLAS Collaboration using its proton-proton (\pp) collision data set collected at $\sqrt{s}=13\TeV$, yielding $\VcbAbs=[5.0^{+1.1}_{-1.4}]\times 10^{-2}$~\cite{atlascollaboration2026measurementvcbelementckm} and demonstrating the experimental viability of this approach.

In this paper, we present the most sensitive direct measurement to date of \VcbAbs using \pp collision data at $\sqrt{s}=13\TeV$.
The analysis builds upon and refines the strategies proposed in Refs.~\cite{Harrison2019,Choi2021}.
The measurement is performed using \ttbar events collected by the CMS experiment.
At the LHC, \ttbar production serves as a copious source of on-shell \PW bosons.
We target the lepton$+$jets final state, where one \PW boson decays leptonically to trigger the event and suppress backgrounds, while the other decays hadronically, providing the signal probe.
In hadronic \PW boson decays, the leading-order partial width to each quark-pair final state is proportional to the squared magnitude of the corresponding CKM matrix element, with corrections from finite quark masses and higher-order QCD and electroweak effects.
The measurement determines \VcbAbs from the relative rate of \Wcb to the other hadronic \PW boson decay modes, primarily \Wud and \Wcs, with smaller contributions from \Wus and \Wcd.
The possible \Wub contribution is proportional to $\abs{V_{\PQu\PQb}}^2$ and is neglected because $\abs{V_{\PQu\PQb}}$ is roughly an order of magnitude smaller than \VcbAbs, making its expected rate much smaller than the experimental sensitivity.
To identify the rare signal amid the other hadronic \PW boson decay backgrounds, we employ a multivariate analysis strategy.
This includes a boosted decision tree (BDT) for the reconstruction of the \ttbar system and a deep neural network discriminant that exploits heavy-flavor (HF) jet properties to isolate the signal.

The paper is structured as follows.
Section~\ref{sec:detector} briefly describes the CMS detector.
The data and simulated samples used in the analysis are detailed in Section~\ref{sec:samples}.
Section~\ref{sec:event_selection_and_ttbar_reconstruction} presents the event selection criteria and the reconstruction of the \ttbar system.
The construction of the multivariate discriminant used to extract the signal is described in Section~\ref{sec:template_score}, followed by a description of the dilepton control region (DL CR) used to constrain the HF background in Section~\ref{sec:cr_dl}.
Systematic uncertainties are discussed in Section~\ref{sec:systematics}.
Finally, the statistical analysis and the results of the measurement are presented in Section~\ref{sec:results}, and a summary is given in Section~\ref{sec:summary}.
Tabulated results are provided in the HEPData record for this analysis~\cite{hepdata}.

\section{The CMS detector}\label{sec:detector}
The CMS apparatus~\cite{CMS:2008xjf,CMS:2023gfb} is a multipurpose, nearly hermetic detector, designed to trigger on~\cite{CMS:2020cmk,CMS:2016ngn,CMS:2024aqx} and identify electrons, muons, photons, and charged and neutral hadrons~\cite{CMS:2020uim,CMS:2018rym,CMS:2014pgm}.
Its central feature is a superconducting solenoid of 6\unit{m} internal diameter, providing a magnetic field of 3.8\unit{T}.
Within the solenoid volume are a silicon pixel and strip tracker, a lead tungstate crystal electromagnetic calorimeter (ECAL), and a brass and scintillator hadron calorimeter (HCAL), each composed of a barrel and two endcap sections.
Forward calorimeters extend the pseudorapidity ($\eta$) coverage provided by the barrel and endcap detectors.
Muons are reconstructed using gas-ionization detectors interleaved with the layers of the steel flux-return yoke outside the solenoid.
More detailed descriptions of the CMS detector, together with a definition of the coordinate system used and the relevant kinematic variables, can be found in Refs.~\cite{CMS:2008xjf,CMS:2023gfb}.

Events of interest are selected using a two-tiered trigger system.
The first level, composed of custom hardware processors, uses information from the calorimeters and muon detectors to select events at a rate of around 100\unit{kHz} within a fixed latency of 4\mus~\cite{CMS:2020cmk}.
The second level, known as the high-level trigger, consists of a farm of processors running a version of the full event reconstruction software optimized for fast processing, and reduces the event rate to a few kHz before data storage~\cite{CMS:2016ngn,CMS:2024aqx}.

The global event reconstruction, also called particle-flow (PF) event reconstruction~\cite{CMS:2017yfk}, aims to reconstruct and identify each individual particle in an event, with an optimized combination of all subdetector information.
In this process, the identification of the particle type (photon, electron, muon, charged hadron, neutral hadron) plays an important role in the determination of the particle direction and energy.
Photons are identified as ECAL energy clusters not linked to the extrapolation of any charged-particle trajectory to the ECAL.
Electrons are identified as a primary charged-particle track and potentially many ECAL energy clusters corresponding to this track extrapolation to the ECAL and to possible bremsstrahlung photons emitted along the way through the tracker material.
Muons are identified as tracks in the central tracker consistent with either a track or several hits in the muon system, and associated with calorimeter deposits compatible with the muon hypothesis.
Charged hadrons are identified as charged-particle tracks neither identified as electrons, nor as muons.
Finally, neutral hadrons are identified as HCAL energy clusters not linked to any charged-hadron trajectory, or as a combined ECAL and HCAL energy excess with respect to the expected charged-hadron energy deposit.

The energy of photons is obtained from the ECAL measurement.
The energy of electrons is determined from a combination of the track momentum at the main interaction vertex, the corresponding ECAL cluster energy, and the energy sum of all bremsstrahlung photons attached to the track.
The energy of muons is obtained from the corresponding track momentum.
The energy of charged hadrons is determined from a combination of the track momentum and the corresponding ECAL and HCAL energies, corrected for the response function of the calorimeters to hadronic showers.
Finally, the energy of neutral hadrons is obtained from the corresponding corrected ECAL and HCAL energies.

For each event, hadronic jets are clustered from these reconstructed particles using the infrared- and collinear-safe anti-\kt algorithm~\cite{Cacciari:2008gp, Cacciari:2011ma} with a distance parameter of 0.4.
Jet momentum is determined as the vectorial sum of all particle momenta in the jet, and is found from simulation to be, on average, within 5 to 10\% of the true momentum over the whole transverse momentum (\pt) spectrum and detector acceptance.
Additional \pp interactions within the same or nearby bunch crossings (pileup) can contribute extra tracks and calorimetric energy depositions to the jet momentum.
To mitigate this effect, charged particles identified to be originating from pileup vertices are discarded and an offset correction is applied to correct for remaining contributions~\cite{CMS:2020ebo}.
Jet energy corrections are derived from simulation to bring the measured response of jets to that of particle level jets on average.
In situ measurements of the momentum balance in dijet, $\text{photon}{+}\text{jet}$, $\PZ{+}\text{jet}$, and multijet events are used to account for any residual differences in the jet energy scale (JES) between data and simulation~\cite{CMS:2016lmd}.
The jet energy resolution (JER) amounts typically to 15--20\% at 30\GeV, 10\% at 100\GeV, and 5\% at 1\TeV~\cite{CMS:2016lmd}.
Additional selection criteria are applied to each jet to remove jets potentially dominated by anomalous contributions from various subdetector components or reconstruction failures~\cite{CMS:2020ebo}.

An additional pileup mitigation algorithm, based on a BDT, is applied to jets with $\pt<50\GeV$.
Performance studies using $\PZ{+}\text{jet}$ events with central jets ($\abs{\eta}<2.5$ and $\pt<50\GeV$) demonstrate an identification efficiency of 99\% for jets originating from the primary \pp interaction and a misidentification rate of 10\% for pileup jets~\cite{CMS:2020ebo}.

To identify (tag) jets originating from \PQb quarks (\PQb jets), information about the energy of particles, tracks, and vertices inside the jet is fed into the \textsc{DeepJet} algorithm~\cite{BTV-16-002,Bols:2020bkb,CMS-DP-2023-005}.
The \PQb tagging discriminant output of the \textsc{DeepJet} algorithm, denoted as the \bvsc score in the following, ranges from zero to one, with larger values indicating a more \PQb-like jet.
The loose, medium, and tight working points (WPs) correspond to misidentification rates of approximately 10, 1, and 0.1\% for light jets, defined here as jets originating from light-flavor quarks or gluons.
The medium WP, which is used for the event selection, provides a \PQb jet tagging efficiency of approximately 70\% and a \PQc jet misidentification rate of roughly 10\%.

The missing transverse momentum vector \ptvecmiss is computed as the negative vector sum of the transverse momenta of all the PF candidates in an event, and its magnitude is denoted as \ptmiss~\cite{CMS:2019ctu}.
The \ptvecmiss is modified to account for corrections to the energy scale of the reconstructed jets in the event.
The pileup-per-particle identification algorithm~\cite{Bertolini:2014bba} is applied to reduce the pileup dependence of the \ptvecmiss observable.
The \ptvecmiss is computed from the PF candidates weighted by their probability to originate from the primary interaction vertex~\cite{CMS:2019ctu}.

\section{Data and simulated samples}\label{sec:samples}

Data sets were collected with the CMS detector from \pp collisions at $\sqrt{s}=13\TeV$ during 2016--2018, corresponding to an integrated luminosity of 138\fbinv.
The data used in this measurement were recorded using triggers that required at least one electron or muon candidate.
The final state of both the background and signal processes contains exactly one lepton.
To suppress nonprompt or misidentified leptons, an isolation requirement is applied at the trigger level.
Here, nonprompt leptons are genuine leptons not produced promptly in the hard-scattering process, whereas misidentified leptons are nonleptonic objects reconstructed as leptons.
The trigger \pt thresholds are 27, 32, and 32\GeV for electrons and 24, 27, and 24\GeV for muons in 2016, 2017, and 2018, respectively.

The \ttbar process is of paramount importance in this measurement, as it constitutes both the signal and the most significant background.
The inclusive \ttbar samples are simulated using the \POWHEG~(v2) generator~\cite{Nason:2004rx,Frixione:2007vw,Alioli:2010xd} at next-to-leading order (NLO) in QCD with the five-flavor scheme (5FS), using the \textsc{hvq} package~\cite{Frixione:2007nw}.
The \PQt quark mass is set to $m_{\PQt}=172.5\GeV$, with the \hdamp parameter fixed at $1.379m_{\PQt}$ to regulate high-energy emissions.
Unless specified otherwise, the NNPDF3.1 next-to-next-to-leading order (NNLO) parton distribution function (PDF) set is consistently used for all simulated processes~\cite{Ball:2017nwa,BALL2012153}.
The inclusive \ttbar samples are normalized to a cross section of 833.9\unit{pb}, calculated at NNLO in QCD with next-to-next-to-leading logarithmic resummation using \TOPpp 2.0~\cite{Czakon_2014}.
Parton showering (PS), hadronization, and the underlying event are simulated using \PYTHIA~(v8.240)~\cite{SJOSTRAND2015159} with the CP5 tune~\cite{Sirunyan2020} for both the signal and background samples.
In this measurement, the lepton$+$jets and dilepton decay channels of the \ttbar process are considered, whereas the fully hadronic decay mode is neglected as its contribution to the events passing the selection criteria described in Section~\ref{sec:event_selection_and_ttbar_reconstruction} is found to be negligible.

A key aspect of the simulation strategy concerns the treatment of hadronic \PW boson decays.
In the \ttbar background samples, quark mixing is modeled considering only the Cabibbo angle, which excludes the \Wcb decay from the background simulation.
To produce signal samples that are kinematically consistent with this background, we force the \Wcb branching fraction to 100\%, while maintaining the same generator settings and tuning as those used for the \ttbar background simulation.
This ensures that any observed differences in the measurement are not due to inconsistencies in the underlying simulation frameworks, allowing for a reliable extraction of \VcbAbs.

For the nominal modeling of \ttbar events with additional \PQb-flavored jets, a dedicated NLO \ttbb prediction is used.
These samples are generated at NLO in QCD using the four-flavor scheme (4FS) with \POWHEGBOX~\cite{Jezo:2015aia,Jezo:2018yaf} and \OPENLOOPS~\cite{Buccioni:2019sur}, explicitly incorporating additional \PQb quarks at the matrix element (ME) level.
The inclusive \ttbar prediction, generated at NLO in QCD, is retained as an alternative description for validation studies; in this description, additional \PQb-flavored jets arise from the PS.
Double-parton scattering (DPS) contributions, which comprise approximately 15\% of the \ttbb cross section, are not covered by the dedicated NLO \ttbb samples and are modeled separately.
The corresponding DPS samples contain \ttbar production generated at NLO in QCD using \POWHEG, while the additional \bbbar production is simulated at leading order (LO) using \PYTHIA with the \texttt{SecondHard} option.

Simulated \ttbar events are categorized according to their additional HF content.
The flavor content is determined at the particle level using an HF-hadron matching procedure.
Generator-level jets used in this categorization are required to satisfy $\pt > 20\GeV$ and $\abs{\eta} < 2.4$.
The \PQb- and \PQc-flavored hadrons are associated with these jets using ghost clustering, in which their momenta are rescaled to infinitesimal values before clustering so that the association does not affect the jet momenta~\cite{Cacciari_2008}.
The ancestry of these hadrons is then traced to identify the origin of the additional HF jets.
Events with additional \PQb-flavored jets are split into three categories: $\ttbar{+}\PQb$, containing exactly one additional \PQb jet with exactly one \PQb-flavored hadron; $\ttbar{+}2\PQb$, containing exactly one additional \PQb jet with two \PQb-flavored hadrons; and $\ttbar{+}\PQb\PQb$, containing exactly two additional \PQb jets.
The $\ttbar{+}\PQb$ component predominantly originates from events in which one of the two additional \PQb jets is outside the jet acceptance.
Events with additional \PQc-flavored jets and no additional \PQb-flavored jets are classified following the same principle as the additional-\PQb categories, into $\ttbar{+}\PQc$ and $\ttbar{+}\PQc\PQc$.
Events containing both additional \PQb- and \PQc-flavored jets are assigned to the additional-\PQb categories.
Remaining events, which do not meet these criteria, are classified as \ttj.
Events containing at least one additional \PQb- or \PQc-flavored jet are collectively referred to as \tthf events.

The single \PQt quark production processes constitute a subdominant background to the \ttbar signal.
The production processes in the $t$ channel and in association with a \PW boson (\tW) are simulated using the \POWHEG generator~\cite{powheg:singletop_st,powheg:singletop_tw}.
In contrast, the $s$-channel single \PQt quark production is simulated at NLO accuracy with \MGvATNLO~(v2.6.5)~\cite{Alwall:2014hca}.
The $s$- and $t$-channel single \PQt quark samples are simulated in the 4FS, whereas the \tW production is simulated in the 5FS.
The diagram removal scheme is used to avoid overlap between \tW and \ttbar production at NLO, where the two processes can interfere~\cite{Frixione:2008,powheg:singletop_tw}.

Minor background contributions originate from several processes, with the most significant among them being the production of \PW and \PZ/\GammaStar bosons in association with additional jets (\Vjets).
These \Vjets samples are simulated using \MGvATNLO at LO accuracy, with up to four additional partons in the ME calculation, and the MLM merging prescription is applied~\cite{Alwall_2007}.
The next contribution comes from \ttbar production in association with additional bosons (\PW, \PZ, and \PH; collectively denoted as \ttX).
The \ttX processes are simulated with \MGvATNLO at NLO accuracy.
For samples generated at NLO, jets described by the ME calculations are consistently combined with those produced by the PS using the \textsc{FxFx} merging prescription~\cite{Frederix:2012ps}.
However, the merging scheme is not employed for the \ttZ process, as additional partons are not included at the ME level.
Finally, diboson production processes (\WW, \WZ, and \ZZ; collectively \VV) are simulated at LO in QCD using the \PYTHIA event generator.

The contribution from QCD multijet events in which a nonprompt or misidentified lepton satisfies the lepton selection is estimated using data.
Inverted lepton selections define QCD-enriched samples orthogonal to the nominal lepton requirements.
With an inverted \ptmiss requirement, transfer factors are derived from the ratio of yields satisfying the nominal and inverted lepton selections, after subtraction of the simulated non-QCD contributions.
The factors are determined as functions of the lepton \pt separately for each data-taking era and are applied to QCD-enriched samples satisfying the nominal \ptmiss requirement, again after subtraction of the simulated non-QCD contributions, to predict the QCD multijet background in the nominal selection.
The selections and \ptmiss thresholds are specified in Section~\ref{subsec:event_selection}.
The resulting QCD multijet contribution after the baseline selection is small compared to the \ttbar background.

For all simulated samples, pileup interactions are overlaid on each simulated hard-scattering event.
The resulting distribution of the number of pileup interactions is reweighted to match the expectation derived from the measured instantaneous luminosity distribution.
The detector simulation is performed using the \GEANTfour toolkit~\cite{Agostinelli:2002hh}.

\section{\texorpdfstring{Event selection and \ttbar reconstruction}{Event selection and ttbar reconstruction}}\label{sec:event_selection_and_ttbar_reconstruction}

\subsection{Event selection}\label{subsec:event_selection}
The event selection categorizes events into two distinct regions based on the lepton multiplicity: a single-lepton (SL) region targeting lepton$+$jets \ttbar decays and a DL CR dominated by dileptonic \tthf events.
In both regions, standard data quality filters reject events affected by detector noise or beam-induced backgrounds, and events containing jets in detector regions known to have anomalous response are vetoed~\cite{CMS:2019ctu}.
The selection requires at least four jets with $\pt>25\GeV$ and $\abs{\eta}<2.4$, of which at least two must be \PQb tagged.

The SL region requires exactly one isolated lepton satisfying tight identification criteria.
Electrons are selected using a multivariate identification method defined in Ref.~\cite{CMS:2020uim}, with the WP optimized to provide a selection efficiency of 80\%, and are required to have $\pt>30\GeV$ in 2016 and $\pt>35\GeV$ in 2017 and 2018, within $\abs{\eta}<2.5$.
Muons are required to meet both tight cut-based identification and isolation requirements as specified in Ref.~\cite{CMS:2018rym}, with $\pt>26\GeV$ in 2016 and 2018 and $\pt>30\GeV$ in 2017, and $\abs{\eta}<2.4$.
Events containing additional leptons satisfying looser selection criteria are rejected to suppress dileptonic \ttbar backgrounds.
The \ptmiss is required to exceed 25\GeV.

For the QCD multijet estimate, the inverted lepton selections require loose-but-not-tight electron identification or a nonisolated muon instead of the nominal lepton requirements.
Transfer factors are measured using events with $\ptmiss<25\GeV$ and applied to the corresponding QCD-enriched samples with $\ptmiss>25\GeV$.

The DL CR is introduced to extract the normalization of \tthf processes and to constrain the associated systematic uncertainties.
The lepton identification and kinematic requirements are the same as those in the SL region, except that exactly two tight isolated leptons with opposite-sign charges are required.
Events containing additional leptons satisfying looser selection criteria are rejected.
Due to the presence of two neutrinos in the final state, the minimum \ptmiss threshold is raised to 40\GeV.
In all dilepton channels, the dilepton invariant mass is required to satisfy $m_{\Pl\Pl}>15\GeV$ to suppress low-mass resonances and backgrounds with nonprompt leptons.
For same-flavor dilepton events, those with $m_{\Pl\Pl}$ within 15\GeV of the \PZ boson mass ($m_{\PZ}$)~\cite{ParticleDataGroup:2022} are rejected to suppress the Drell--Yan background.

The event selection is summarized in Table~\ref{tab:event_selection}.

\begin{table}[htbp]
  \centering
  \topcaption{Summary of event selection.}
  \label{tab:event_selection}
  \begin{tabular}{lll}
      Variable                                     & SL channel       & DL channel       \\
      \hline
      Data quality filters                         & standard         & standard         \\
      Problematic jet regions                      & vetoed           & vetoed           \\
      Number of jets                               & $\geq$4          & $\geq$4          \\
      Number of \PQb-tagged jets                   & $\geq$2          & $\geq$2          \\
      Number of tight leptons                      & 1                & 2                \\
      Additional lepton veto                       & yes              & yes              \\
      Charge of lepton pair                        & \NA              & opposite sign    \\
      $\abs{m_{\Pl\Pl} - m_{\PZ}}$ (same-flavor) & \NA              & $>$15\GeV \\
      $m_{\Pl\Pl}$                               & \NA              & $>$15\GeV \\
      \ptmiss                                      & $>$25\GeV & $>$40\GeV \\
  \end{tabular}
\end{table}

\subsection{\texorpdfstring{Reconstruction of \ttbar system}{Reconstruction of ttbar system}}\label{subsec:ttbar_reconstruction}
The reconstruction of the lepton$+$jets \ttbar system is essential for this measurement, as both the signal and the primary background originate from \ttbar production.
The dominant backgrounds are \ttbar events in which the hadronically decaying \PW boson produces $\PQu\PQd$ or $\PQc\PQs$ quark pairs, including the corresponding Cabibbo-suppressed $\PQu\PQs$ and $\PQc\PQd$ modes.
These processes pose a particular challenge when accompanied by additional \PQb quark activity, which can mimic the final-state topology of the signal.
To distinguish the signal from these backgrounds, an accurate assignment of the final-state jets to their respective parent partons is required.

This assignment involves mapping four reconstructed jets to the corresponding quarks in the \ttbar decay chain.
We define these four jets based on their production branch: \jhadtb and \jleptb represent the \PQb jet candidates produced directly from the hadronic and leptonic \PQt quark decays, respectively.
The two jets resulting from the hadronic \PW boson decay are denoted as \jwu and \jwd, corresponding to the up-type (\PQu or \PQc) and down-type (\PQd, \PQs, or \PQb) quark candidates.
For example, in the signal process \Wcb, \jwu and \jwd specifically represent the \PQc and \PQb quark candidates from the \PW boson decay, respectively.

The \ttbar reconstruction is performed using two sequential BDTs based on XGBoost~\cite{Chen:2016:XST:2939672.2939785}.
The first BDT selects the overall jet-parton configuration, including the \jwu and \jwd candidates, while the second BDT resolves their up--down assignment.
This two-step approach is designed to reduce the misassignment of jets from additional HF radiation in \tthf events as hadronic \PW boson decay products.
Both BDTs are trained on simulated signal samples to differentiate between correct and incorrect assignments.

The first BDT considers all possible assignments of the reconstructed jets to \jhadtb, \jleptb, and a pair of \PW boson daughter-jet candidates.
It also considers the possible solutions for the neutrino longitudinal momentum (\pzN) obtained from the \PW boson mass constraint, taking \ptvecmiss as the neutrino transverse momentum.
If the solutions are complex, only their real component is used.
For each event, the BDT is evaluated separately for every combination of a jet-parton assignment and a \pzN solution.
The combination yielding the highest BDT score is selected.
The input variables include known kinematic constraints, such as $m_{\PW}$ and $m_{\PQt}$, various angular distributions, and the expected momentum balance between the \PQt and \PAQt quarks.
The \bvsc scores of the \textsc{DeepJet} algorithm for the jets assigned as candidates for \jhadtb and \jleptb are also included.
In contrast, the \bvsc scores of the two \PW boson daughter-jet candidates are intentionally omitted.
Because these scores provide strong flavor discrimination, including them would cause the BDT to preferentially assign jets from additional HF radiation in \tthf events as hadronic \PW boson decay products.

The second BDT resolves the ambiguity between the two selected \PW boson daughter-jet candidates by assigning them to \jwu and \jwd.
Without the \bvsc scores of these candidates, the remaining kinematic inputs are largely symmetric under their exchange.
A small residual discrimination remains through the flavor-dependent \PQb jet energy regression~\cite{Sirunyan:2020qvi}, but it is insufficient to resolve the ambiguity reliably.
The second BDT therefore evaluates the two possible assignments using the \bvsc scores of the two candidates.
The assignment yielding the higher BDT score is selected, and the corresponding score is denoted by \Dreco.

For events in which the two \PW boson daughter jets are correctly selected, the performance of this ordering is assessed through the up--down label-exchange rate at the generator level.
Before applying the dedicated assignment BDT, the exchange rates are approximately 46\%, 46\%, and 41\% for \Wud, \Wcs, and \Wcb decays, respectively.
For the signal, the high \bvsc score of the \PQb jet reduces the exchange rate to approximately 13\%.
For \Wcs decays, the \PQc jet tends to have a higher \bvsc score than the \PQs jet because \PQc-flavored hadron decays can produce displaced tracks and secondary vertices.
The \PQc jet is thus assigned as the down-type candidate in about 68\% of events.
For \Wud decays, the two light-quark jets are not distinguishable with the available flavor-tagging information, and the exchange rate remains approximately 47\%, as expected.

For the BDTs used to select the jet-parton mapping, the response distributions in the training and validation samples are compared using Kolmogorov--Smirnov tests~\cite{Hodges:1958}, and no evidence of overtraining is observed.

To evaluate the overall performance of the \ttbar reconstruction, the selected jet-parton assignments are categorized into four distinct classes based on the generator-level information:
\begin{itemize}
  \item \PW boson daughters found: the hadronically decaying \PW boson daughters are accurately matched.
  \item Combinatorial \PW boson daughter misassignments: all hard-process jets pass the baseline selection, but the \PW boson daughters are incorrectly identified among them.
  \item Out-of-acceptance losses: at least one hard-process jet fails the baseline selection, leading to an inevitable misassignment.
  \item Radiative HF contamination: a jet originating from additional HF radiation is erroneously assigned as a \PW boson daughter.
\end{itemize}

Approximately 50\% of \ttbar events containing four hard-process jets pass the baseline selection criteria described in Section~\ref{subsec:event_selection}.
For these selected events, the efficiency of achieving a correct assignment (the ``\PW~found'' category) depends on the specific decay mode.
The identification efficiencies are found to be 86\% for \Wud, 81\% for \Wcs, and 67\% for the signal \Wcb decay.
The \tthf process, however, introduces additional complexity; depending on the \PW decay mode and radiation activity, approximately 6--10\% of these events fall into the HF contamination category, where an HF jet from radiation is incorrectly identified as a \PW boson decay product.

\begin{figure}[htbp]
  \centering
  \includegraphics[width=1.0\textwidth]{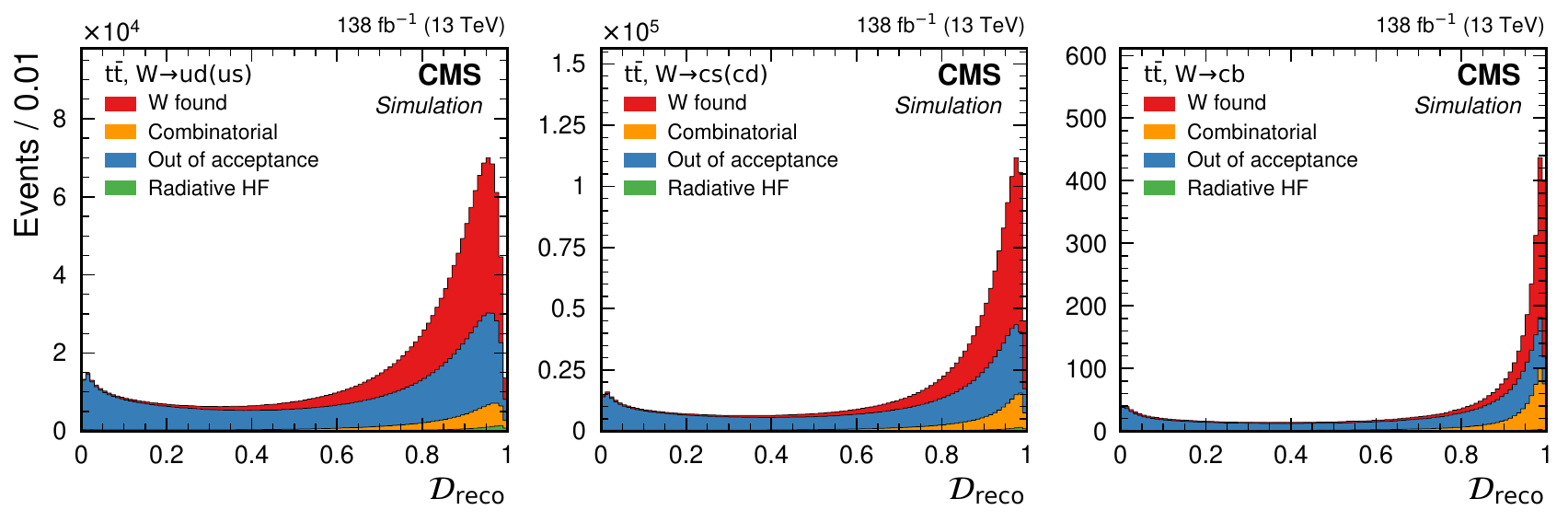}
  \caption{
    Distributions of the \Dreco evaluated on simulated lepton$+$jets \ttbar events, categorized by the hadronic \PW boson decay mode: \Wudus (left), \Wcscd (center), and the signal \Wcb (right).
    The stacked histograms represent the categorization of the selected jet-parton assignments based on generator-level information, as defined in the text.
  }\label{fig:reco_bdt_score}
\end{figure}

Figure~\ref{fig:reco_bdt_score} shows the \Dreco distribution for the lepton$+$jets \ttbar process according to these defined categories.
The dominant source of reconstruction failure is the loss of at least one hard-process jet outside the detector acceptance.
As shown in Fig.~\ref{fig:reco_bdt_score}, such events predominantly populate the low-\Dreco region.

\section{\texorpdfstring{Construction of the \Dcb discriminant}{Construction of the Dcb discriminant}} \label{sec:template_score}
Following the \ttbar reconstruction, a single multivariate discriminant, denoted as \Dcb, is constructed, and its output is used as a one-dimensional template for the statistical extraction of \VcbAbs.
To maximize the sensitivity to the \Wcb signal while ensuring robustness against dominant backgrounds, the TabNet algorithm~\cite{arik2020tabnetattentiveinterpretabletabular} is employed.
TabNet is a deep neural network architecture specifically optimized for tabular data sets, such as those encountered in high-energy physics.
A key advantage of TabNet is its built-in inductive bias, which enables the network to identify relevant input variables (hereafter referred to as features) for each event.
Sequential attention masks assign weights to these features, allowing the model to focus on the most discriminating properties of the reconstructed \ttbar system.

The TabNet model is trained using lepton$+$jets \ttbar signal and background samples and is optimized for multiclass classification into the seven categories summarized in Table~\ref{tab:definition_classification}.
These classes are defined by three primary criteria: the hadronic \PW boson decay mode, the presence of additional HF jets, and the success of the jet-parton assignment during the \ttbar reconstruction.
This classification scheme allows the network to exploit both flavor information and reconstruction quality when separating the \Wcb signal from the dominant \ttbar backgrounds.

\begin{table}[htbp]
  \centering
  \topcaption{
    Definition of the seven training classes for the \Dcb discriminant.
    The events are categorized into seven distinct classes based on the hadronic \PW boson decay mode, the presence and flavor of additional HF jets, and the success of the jet-parton matching during the \ttbar reconstruction.
    Classes 1 and 2 constitute the targeted signal process, while Classes 3 through 7 represent the various background topologies.
    The additional \PQb-flavored jet class comprises the $\ttbar{+}\PQb$, $\ttbar{+}2\PQb$, and $\ttbar{+}\PQb\PQb$ generator-level categories, whereas the additional \PQc-flavored jet class comprises the $\ttbar{+}\PQc$ and $\ttbar{+}\PQc\PQc$ generator-level categories.
  }\label{tab:definition_classification}
  \begin{tabular}{clll}
    Class & \PW decay mode & Additional HF jets                        & Jet-parton matching \\
    \hline
    1     & \Wcb           & Inclusive                                 & Matched             \\
    2     & \Wcb           & Inclusive                                 & Unmatched           \\
    3     & \Wudus         & Additional \PQb-flavored jets             & Inclusive           \\
    4     & \Wudus         & No additional \PQb-flavored jets          & Inclusive           \\
    5     & \Wcscd         & Additional \PQb-flavored jets             & Inclusive           \\
    6     & \Wcscd         & Additional \PQc-flavored jets             & Inclusive           \\
    7     & \Wcscd         & No additional \PQb- or \PQc-flavored jets & Inclusive           \\
  \end{tabular}
\end{table}

Separate training and output weights are used to maintain stable learning of the rare signal classes and optimize the expected sensitivity of the measurement.
Training with weights proportional to the expected process yields would give the rare signal classes negligible influence on the loss function and could destabilize the training.
The classifier is therefore trained with balanced weights that give signal and background events comparable influence on the loss function.
Because the resulting class composition differs from that expected in the analysis regions, separate class-dependent weights are applied to the classifier outputs after training.
These output weights are chosen to maximize the expected sensitivity while accounting for statistical and dominant systematic uncertainties.
The optimization improves the expected sensitivity metric by approximately 20\% relative to the unweighted discriminant.
The weighting procedure and its optimization are detailed in Appendix~\ref{app:class_weights}.

The \Dcb discriminant is constructed by summing the weighted outputs for the two signal classes (Classes 1 and 2).
It ranges from 0 to 1, with values near unity indicating an event more compatible with the \Wcb signal.

The inputs to the \Dcb discriminant are chosen to exploit the characteristic flavor content of the \Wcb final state while reducing sensitivity to kinematic features unrelated to the hard process.

\begin{itemize}
  \item Jet flavor: The \textsc{DeepJet} \bvsc scores of the jets assigned as \jwu and \jwd candidates are fundamental to the discriminant's performance.
        In \Wcb decays, the genuine \PQb and \PQc jets originate from the hadronically decaying \PW boson.
        While the genuine \PQb jet provides a high \bvsc response, the \PQc jet also exhibits ``\PQb-like'' tagging characteristics that result in moderately high scores.
        This combined signature provides powerful separation against Cabibbo-favored backgrounds: \Wcs events typically contain only a single \PQc jet with a moderate score, and \Wud events lack significant HF tagging responses entirely.

        The \bvsc scores for \jhadtb and \jleptb also help identify reconstruction failures.
        Specifically, this enables the model to recognize cases where a jet from the \PW boson decay is incorrectly assigned as \jhadtb or \jleptb, or vice versa, during the reconstruction process.

  \item Reconstruction fidelity: To identify misreconstructed events, \Dreco and the masses of the hadronically decaying \PW boson and \PQt quark are used to quantify the confidence in the jet-parton assignment.
        Additionally, the jet and \PQb jet multiplicities, along with the scalar sum of the jet \pt values (\HT), are incorporated to characterize the radiation levels and the overall reconstruction complexity.

  \item Suppression of HF radiation: The minimum invariant mass (\minMbb) and angular separation (\minRbb) among all pairs of \PQb-tagged jets in each event are provided as inputs to the TabNet discriminant.
        Additional HF radiation, particularly gluon splitting into a collimated, low-mass \bbbar pair, can produce small values of these observables.
        They therefore provide information about the additional HF activity and reconstruction complexity of the event.

  \item Hard-process kinematic variables: The \pt of \jhadtb, \jwu, \jwd, and \jleptb, along with the \pt of the entire \ttbar system, are provided to exploit the production dynamics.

  \item Data-taking era: An identifier for the specific data-taking year is included to account for the effects caused by variations in detector conditions across different data-taking periods.
\end{itemize}

\begin{figure}[htbp]
  \centering
  \includegraphics[width=0.55\textwidth]{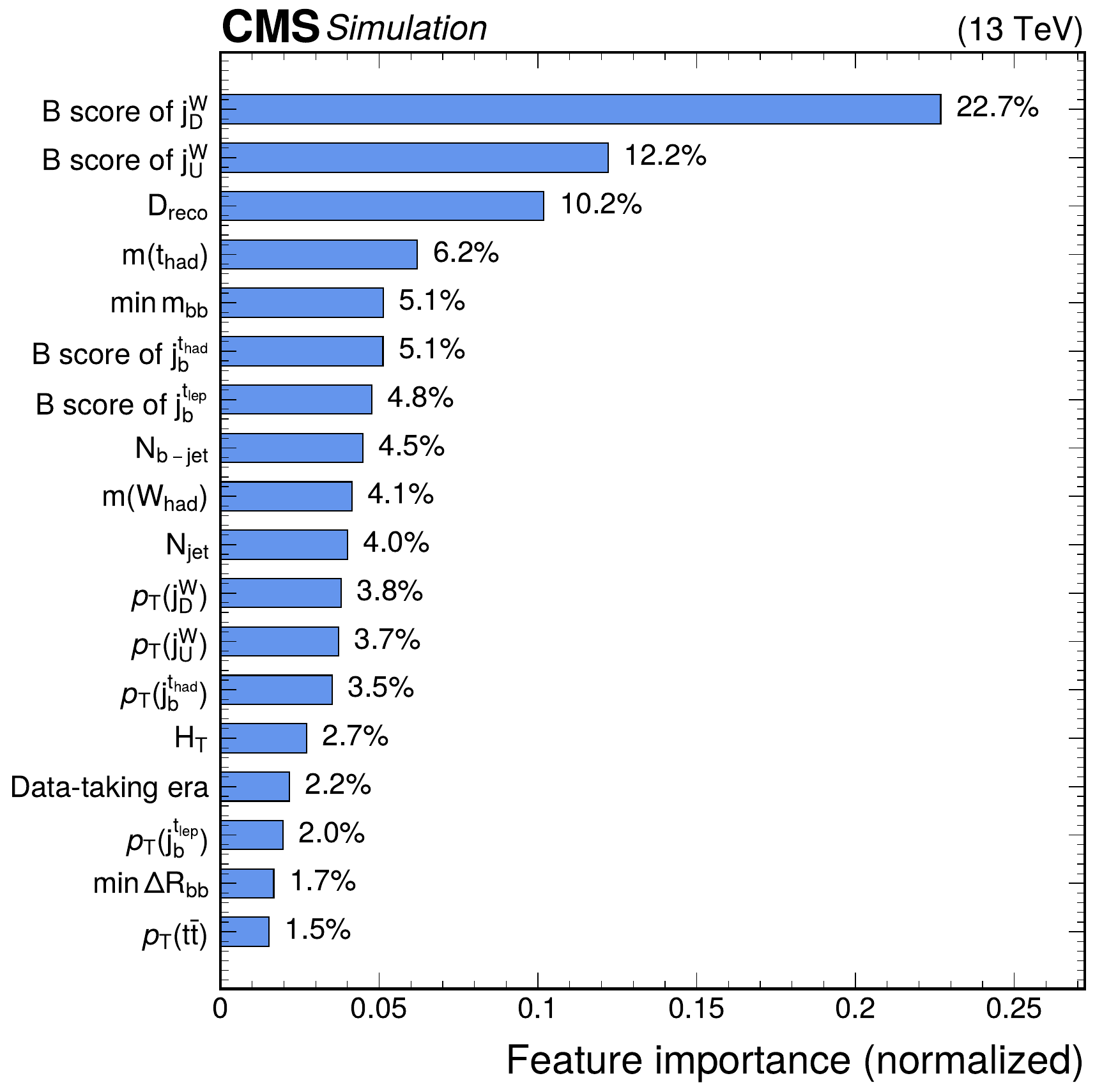}
  \caption{
    Feature importance for the \Dcb discriminant as determined by the TabNet model during training.
  }\label{fig:tabnet_input_importance}
\end{figure}

The global importance of all input features used by the classifier, shown in Fig.~\ref{fig:tabnet_input_importance}, is obtained following the prescription in Ref.~\cite{arik2020tabnetattentiveinterpretabletabular}: the feature-selection masks are weighted by the corresponding decision-step contributions, summed over all decision steps, normalized across the input features, and averaged over the training events.
As expected, the \bvsc scores of the \jwd and \jwu candidates emerge as the two most prominent features for the classification.
Furthermore, the \ttbar reconstruction fidelity score, \Dreco, ranks as the third most important input.
The reconstructed $m_{\PQt}$ and the \minMbb observable rank fourth and fifth, respectively.
The resulting hierarchy is consistent with the a priori expectation that the flavor-tagging information of the hadronic \PW boson decay jets and the quality of the \ttbar reconstruction should be among the most discriminating inputs.
This provides a qualitative validation that the model has learned a physically meaningful classification strategy.

To ensure the robustness of the network and prevent potential bias from overfitting, a 3-fold out-of-fold inference strategy is employed.
In this approach, the data set is randomly partitioned into three independent subsets.
Three separate models are trained, each using two of the subsets for training while reserving the third for evaluation.
This ensures that every event is evaluated by a model whose training sample did not include that event, thereby providing out-of-fold predictions across the entire data set.

\begin{figure}[htbp]
  \centering
  \includegraphics[width=0.55\textwidth]{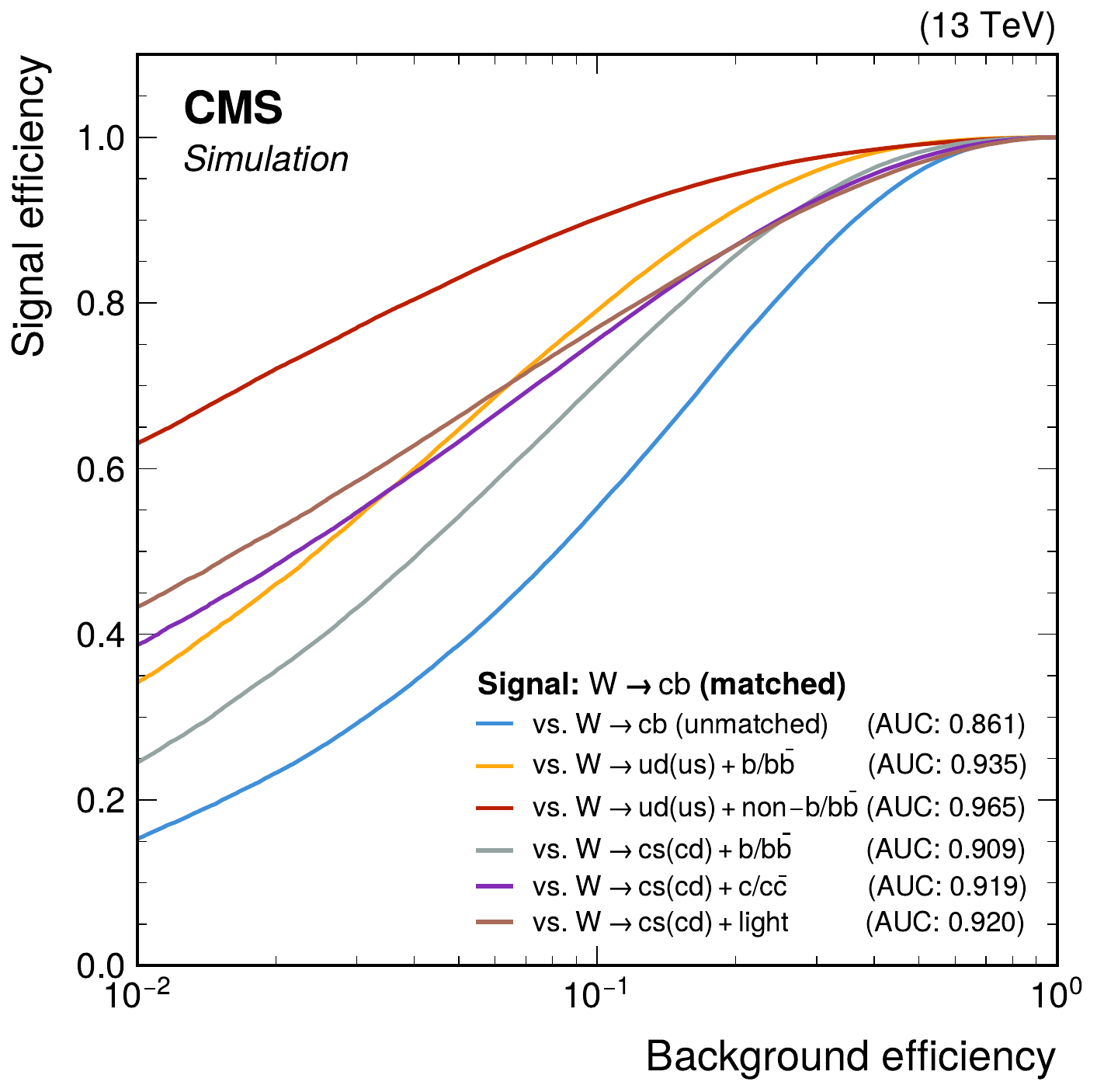}
  \caption{
    Signal efficiency versus background efficiency for separating correctly matched \Wcb signal events from each of the other event classes.
    AUC denotes the area under each curve.
    The definitions of the signal and background classes are given in Table~\ref{tab:definition_classification}.
  }\label{fig:roc}
\end{figure}

Figure~\ref{fig:roc} shows how effectively the \Dcb discriminant separates correctly matched \Wcb signal events from the other classes, which are defined by the hadronic \PW boson decay mode, additional HF jet content, and jet-parton matching.
As anticipated, the Cabibbo-favored \Wcs category with additional \PQb-flavored jets is among the most challenging backgrounds because of its similar HF content to the signal.
The \Wcscd category with additional \PQb-flavored jets (Class 5 in Table~\ref{tab:definition_classification}) is found to have the lowest separation power among the background classes, with an area under the curve (AUC) of 0.909.

For the final fit, only template bins with $\log(S/B)>-2.25$, where $S$ and $B$ denote the expected signal and background yields, are retained.
Bins with lower expected $S/B$ ratios contain negligible signal contributions and do not improve the expected sensitivity.

\begin{figure}[htb]
  \centering
  \includegraphics[width=1.0\textwidth]{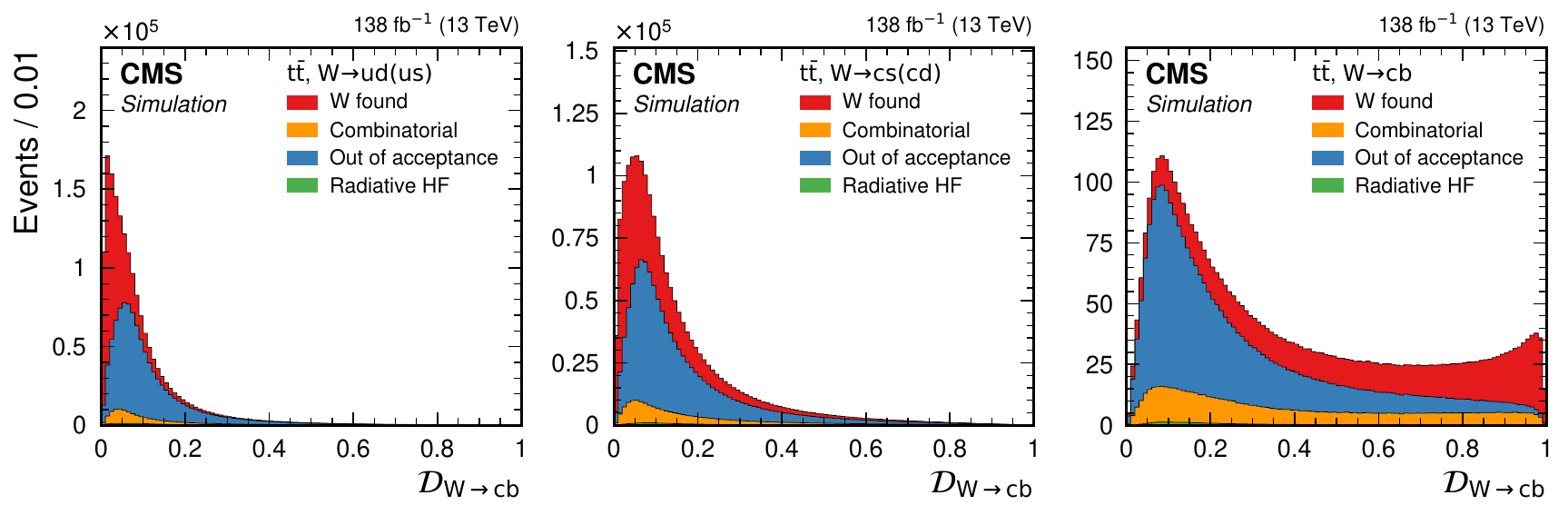}
  \caption{
    Distributions of the \Dcb discriminant for simulated events, categorized by the hadronic \PW boson decay mode: \Wudus (left), \Wcscd (center), and \Wcb (right).
    The stacked histograms represent the categorization of the jet-parton assignments, as explained in Section~\ref{subsec:ttbar_reconstruction}.
  }\label{fig:template_score}
\end{figure}

The composition of the \Dcb discriminant for different hadronic \PW boson decay modes is shown in Fig.~\ref{fig:template_score}.
The three panels illustrate the distributions for \Wudus, \Wcscd, and the signal \Wcb, respectively.
For each channel, the events are categorized according to their reconstruction fidelity, as previously defined in Section~\ref{subsec:ttbar_reconstruction}.
For the \Wcb signal process, the correctly matched events (denoted as ``\PW~found'') are shifted toward higher \Dcb values.
This separation enables a clear identification of the signal.

\begin{figure}[htbp]
  \centering
  \includegraphics[width=0.48\textwidth]{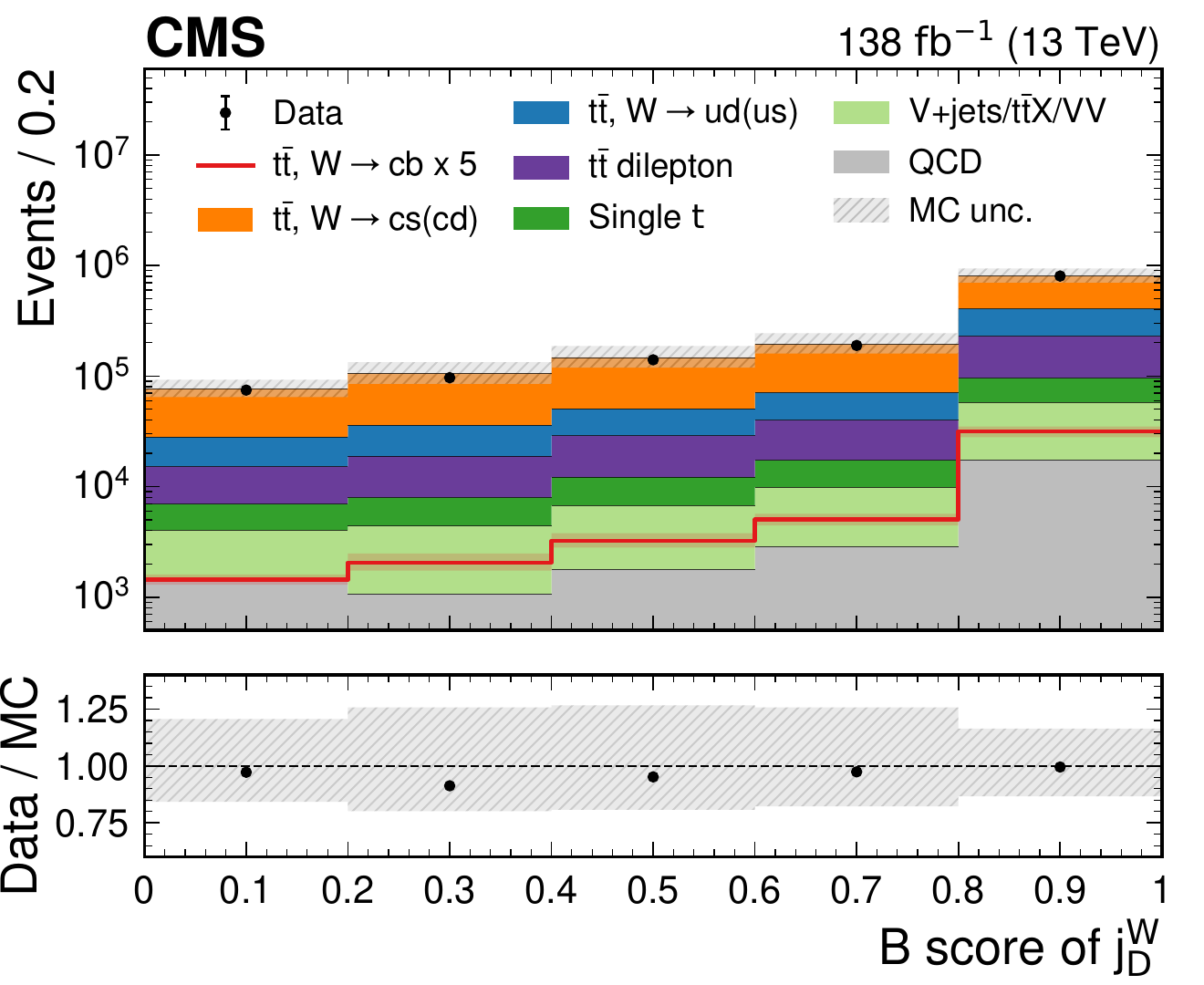}
  \includegraphics[width=0.48\textwidth]{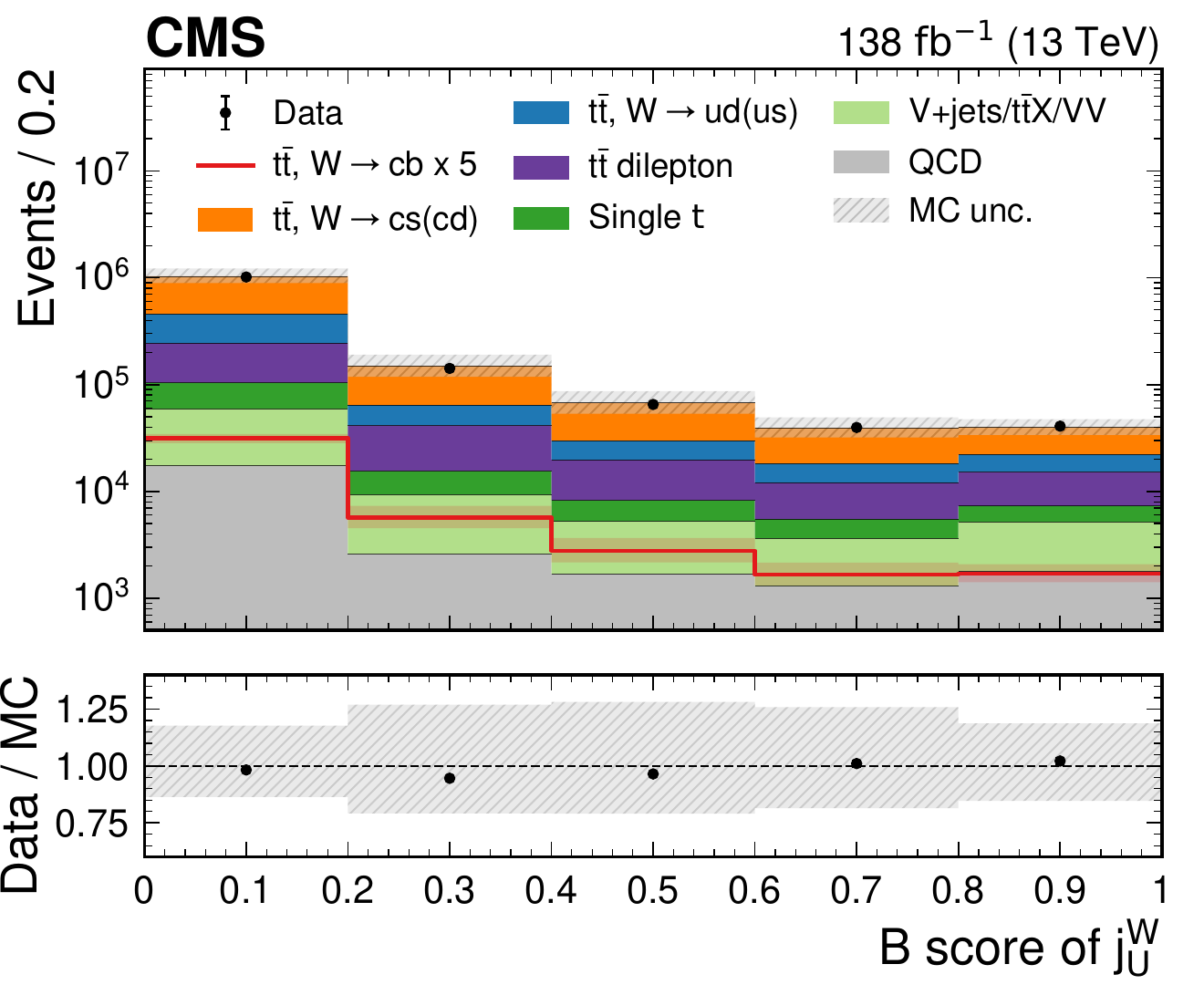}
  \includegraphics[width=0.48\textwidth]{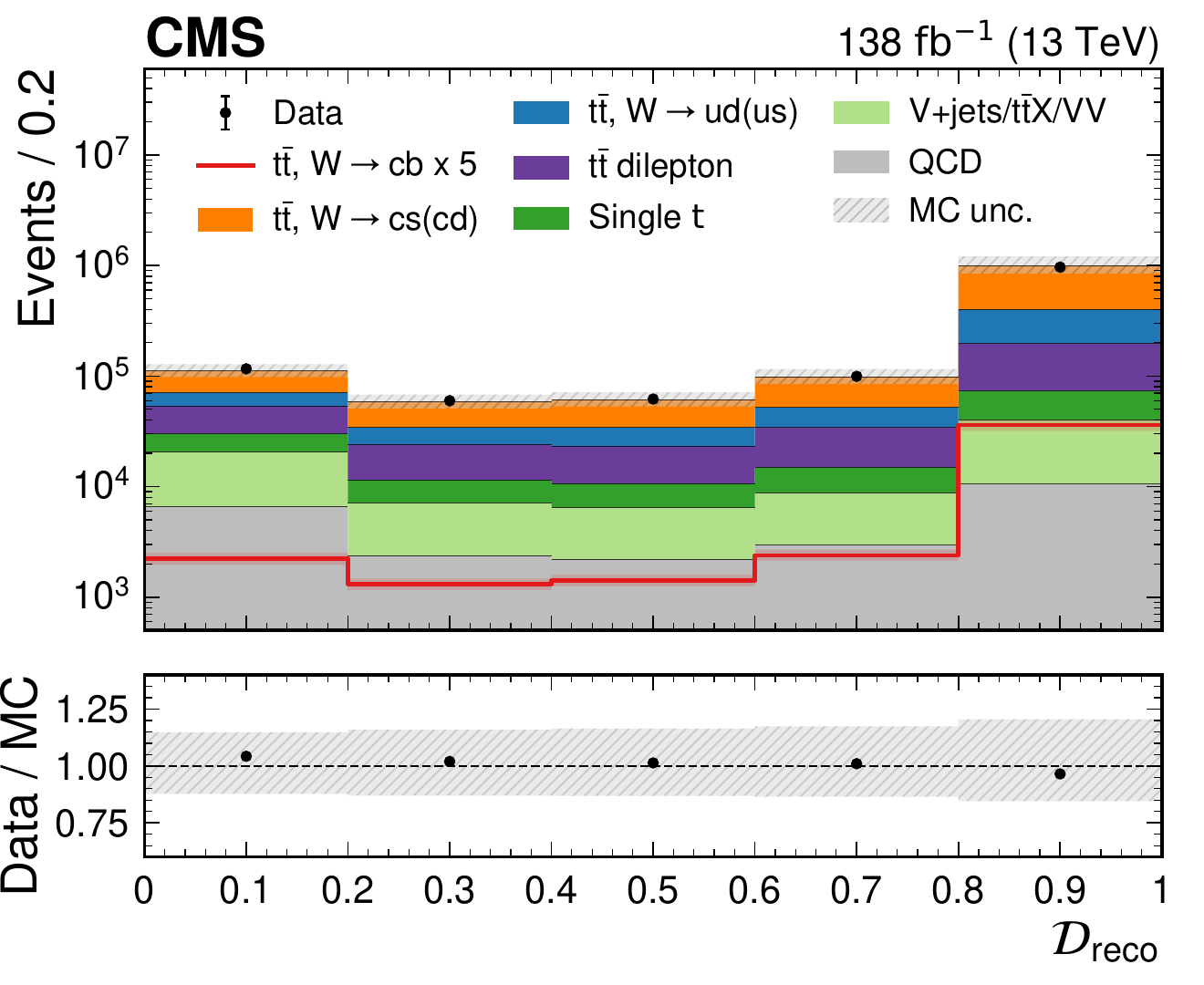}
  \includegraphics[width=0.48\textwidth]{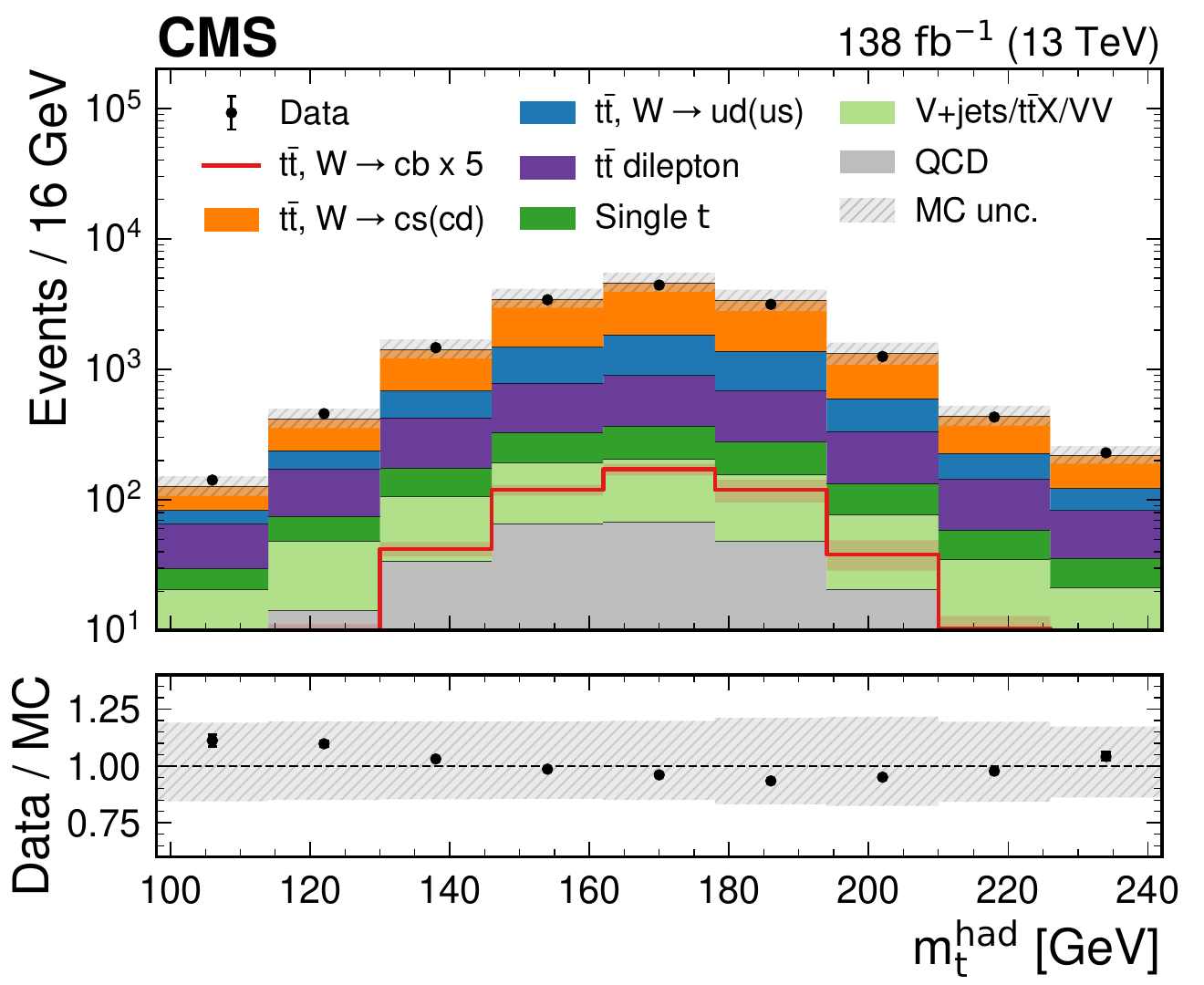}
  \includegraphics[width=0.48\textwidth]{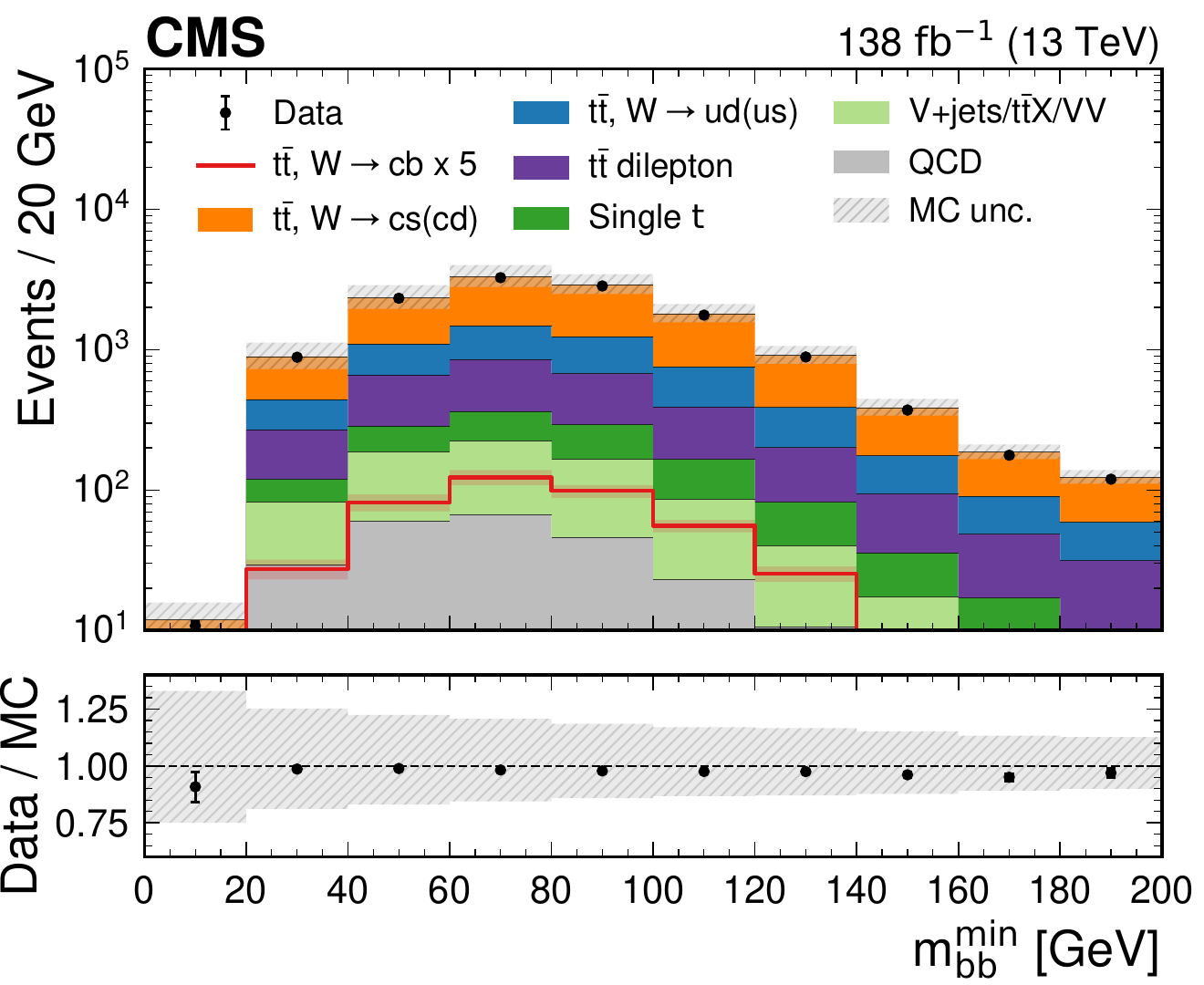}
  \includegraphics[width=0.48\textwidth]{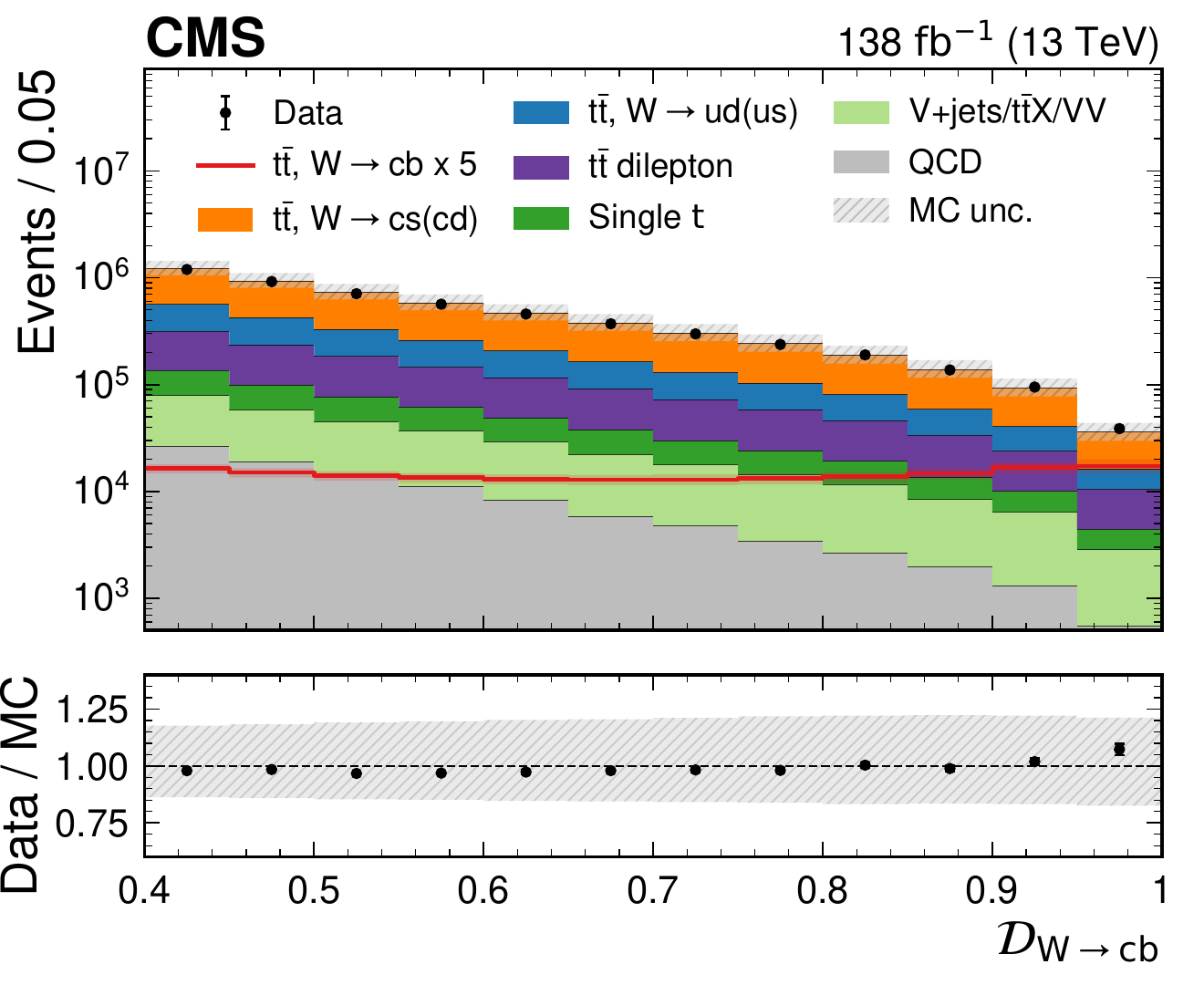}
  \caption{
    Pre-fit distributions in the SL region of the \bvsc scores of the \jwd (upper left) and \jwu (upper right) candidates, the \Dreco discriminant (middle left), the reconstructed $m_{\PQt}$ (middle right), the \minMbb observable (lower left), and the \Dcb discriminant (lower right).
    For visibility, the signal contribution is overlaid at five times the rate corresponding to the PDG reference value of \VcbAbs.
    It is not included in the stacked background prediction.
    Contributions from \tthf are included in the \ttbar categories labeled by the hadronic \PW boson decay mode and in the \ttbar dilepton category.
    The vertical error bars on the data points indicate their statistical uncertainties.
    The uncertainty bands include statistical and systematic components before the likelihood fit.
  }\label{fig:prefit_template}
\end{figure}

Figure~\ref{fig:prefit_template} presents the pre-fit data and simulation distributions of the five highest-ranked input variables and the \Dcb discriminant.
The pre-fit predictions describe the data within the assigned uncertainties, whose sources are discussed in Section~\ref{sec:systematics}, providing a validation of the event modeling before the likelihood fit described in Section~\ref{sec:results}.
The comparatively large pre-fit bands are driven mainly by the \muR variations of the dedicated 4FS \ttbb prediction and by the flavor-tagging calibration uncertainties, including conservative variations of the \PQc jet response.
The systematic variations are correlated across bins, so the band should not be compared directly with the point-to-point fluctuations of the data.

\section{\texorpdfstring{Dilepton control region for \tthf backgrounds}{Dilepton control region for tt+HF backgrounds}}\label{sec:cr_dl}
The \tthf process represents a significant background within the SL region.
Its modeling is challenging because of the multiscale QCD dynamics and the interplay with PS modeling~\cite{Jezo:2018yaf}.
To improve the modeling accuracy and mitigate associated uncertainties, we perform a simultaneous fit incorporating a dedicated DL CR.
This approach uses shared nuisance parameters across regions, allowing the data in the DL CR to provide direct constraints on the background components in the SL region.

The DL CR is specifically designed to be enriched in \tthf events, while signal contamination remains negligible.
The event selection criteria for this region are detailed in Section~\ref{subsec:event_selection}.
When jets are ordered by their \bvsc score, the two leading jets are typically associated with the \PQb quarks from the \PQt quark decays, while the third and fourth jets are more sensitive to additional HF radiation.

The DL CR is constructed using a two-dimensional (2D) distribution defined by the \bvsc scores of the third and fourth jets in the score-ordered jet collection.
For each jet, the score is divided into four intervals: $[0,L)$, $[L,M)$, $[M,T)$, and $[T,1]$, labeled 0, L, M, and T, respectively, where $L$, $M$, and $T$ denote the era-dependent loose, medium, and tight WP thresholds.
Because the jets are ordered by decreasing \bvsc score, the third jet is always in the same or a higher score interval than the fourth, leaving ten populated cells.
These cells are unrolled by increasing the score interval of the third jet and, within each interval, that of the fourth jet.
The statistical fit retains the final six bins, labeled ML, MM, T0, TL, TM, and TT, in that order; the first and second characters of each label refer to the third and fourth jets, respectively.
The four lower-score cells are excluded to make the selected \tthf composition in the DL CR more similar to that in the SL region.
The resulting histogram is referred to as the unrolled 2D \bvsc template.
This binning scheme helps distinguish \ttj events, which are concentrated in lower-score bins, from \tthf events, which tend to populate higher-score bins, and thereby constrains the \tthf background in the simultaneous fit.

\begin{figure}[htbp]
  \centering
  \includegraphics[width=0.6\textwidth]{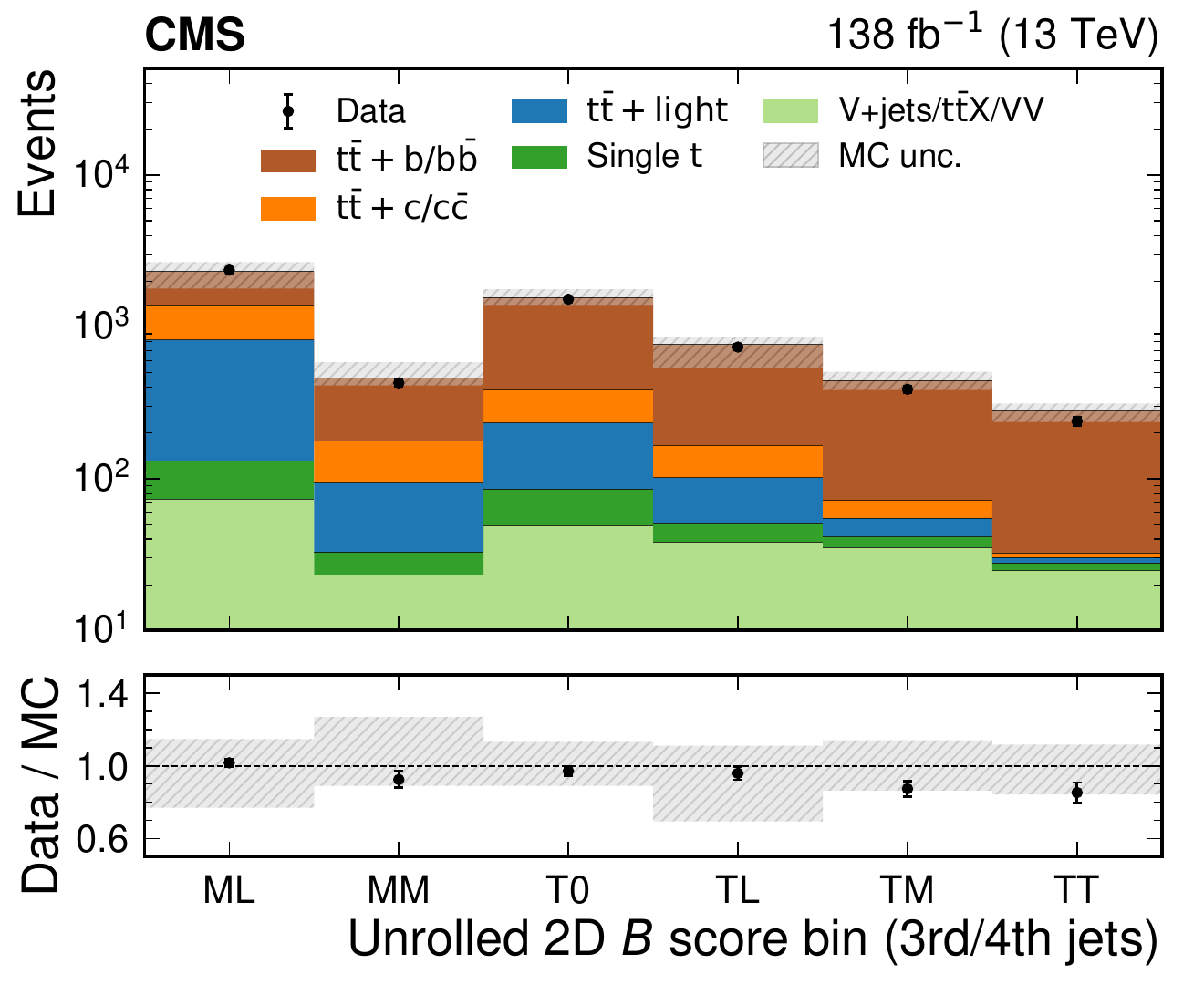}
  \caption{
    Pre-fit distribution of the unrolled 2D \bvsc template in the DL CR.
    The \ttbar contributions are grouped by the flavor content of additional jets; the detailed generator-level categories are defined in Section~\ref{sec:samples}.
    The vertical error bars on the data points indicate their statistical uncertainties.
    The uncertainty band includes statistical and systematic components before the likelihood fit.
  }\label{fig:prefit_dl}
\end{figure}

Figure~\ref{fig:prefit_dl} shows the pre-fit distribution of this template in the DL CR.

\section{Systematic uncertainties}\label{sec:systematics}
Various corrections are applied to the simulation samples.
The uncertainties associated with these corrections serve as sources of systematic uncertainties.
All systematic uncertainties are treated as nuisance parameters in the statistical model.
Rate uncertainties are modeled by multiplicative scale variations following a log-normal distribution.
Shape uncertainties affect the distribution on a bin-by-bin basis and are modeled using polynomial interpolation with Gaussian variations.

\subsection{Modeling and theoretical uncertainties}
The effects of all modeling variations on the process normalization are treated separately and included in the process normalization uncertainties described below, while the corresponding variations are treated only as shape uncertainties.
\begin{itemize}
  \item Renormalization and factorization scales: The uncertainty associated with missing higher-order QCD corrections is evaluated by varying the renormalization (\muR) and factorization (\muF) scales separately at the ME level by factors of 2 and 0.5, with the other scale fixed to its nominal value.
        The \muR and \muF variations are implemented as separate shape uncertainties, correlated across the data-taking years.
        They are treated independently for \ttbar and single \PQt quark production.
        Within the \ttbar prediction, the scale uncertainties for the additional-\PQb components modeled with the \OPENLOOPS samples are decorrelated from those of the remaining \ttbar components.

  \item Additional HF normalization and composition: The nominal modeling of \ttbar events with additional \PQb-flavored jets is based on the dedicated NLO \ttbb samples described in Section~\ref{sec:samples}.
        The total normalization of the combined additional-\PQb component, including $\ttbar{+}\PQb$, $\ttbar{+}2\PQb$, $\ttbar{+}\PQb\PQb$, and the dedicated \ttbb DPS contribution, is allowed to float freely as a single parameter in the fit.
        For the combined additional-\PQc component, consisting of $\ttbar{+}\PQc$ and $\ttbar{+}\PQc\PQc$ events, the normalization is constrained using a scale factor of 1.11 with a 21\% Gaussian prior, based on the central value and precision measured in Ref.~\cite{2021136565}.
        In addition, the relative fractions of the subcomponents within each HF category are varied while preserving the corresponding total additional-\PQb or additional-\PQc rates.
        Gaussian uncertainties of 50\% are assigned to the $\ttbar{+}2\PQb/\ttbar{+}\PQb$, $\ttbar{+}\PQb\PQb/\ttbar{+}\PQb$, and $\ttbar{+}\PQc\PQc/\ttbar{+}\PQc$ ratios.

  \item Modeling of \PQt quark \pt in \ttbar events: The simulated \PQt quark \pt spectrum in \ttbar events is known to be harder than that observed in data~\cite{PhysRevD.97.112003}.
        To correct for this effect, the simulated \ttbar events generated at NLO in QCD are reweighted to match the \pt spectrum at NNLO accuracy in QCD.
        This reweighting is implemented using a neural network-based method~\cite{CMS:2025mlr}.
        The NNLO-reweighted distribution is used for both the nominal prediction and the downward variation, while the unweighted NLO distribution defines the upward variation of this one-sided uncertainty.

  \item Parton shower and event modeling: These uncertainties are evaluated using dedicated variations of the simulated samples:
        \begin{itemize}
          \item Matrix element--parton shower matching: The \POWHEG parameter \hdamp is varied from its nominal value of $1.379m_{\PQt}$~\cite{Sirunyan2020} to $2.305m_{\PQt}$ and $0.8738m_{\PQt}$ for the up and down variations, respectively.
          \item Initial- and final-state radiation: The amounts of initial- and final-state radiation are varied by changing the value of the strong coupling constant \alpS used in the PS at the relevant shower scale.
                Variations in initial-state radiation are treated independently for \ttbar and single \PQt quark production, whereas variations in final-state radiation are treated as correlated.
          \item Fragmentation of \PQb quarks: The uncertainty in the fragmentation of \PQb quarks is evaluated by varying the \pt of generator-matched \PQb jets in simulated \ttbar events by $\pm$1\%.
          \item Color reconnection: Alternative color reconnection models are used to assess the uncertainty associated with the modeling of color flow during hadronization~\cite{Argyropoulos2014,Christiansen2015}.
          \item Underlying event: The parameters of the CP5 tune are varied within their uncertainties~\cite{Sirunyan2020}.
        \end{itemize}
        The resulting variations are propagated to the final templates.

  \item Mass of the \PQt quark: The uncertainty from $m_{\PQt}$ is estimated by comparing the nominal sample with samples where $m_{\PQt}$ is varied by $\pm$1\GeV.
        Due to the limited event counts in the dedicated samples, only the effect of the mass variation on the event-selection acceptance is retained as a rate uncertainty.

  \item Process normalization uncertainties: The simulated samples are normalized to their respective theoretical cross sections, and the corresponding normalization uncertainties are included as rate uncertainties.
        For the inclusive \ttbar process, a combined rate uncertainty of 5.2\% is assigned, accounting for scale, PDF, \alpS and $m_{\PQt}$ uncertainties.
        For non-\ttbar backgrounds, separate normalization uncertainties are assigned according to the expected size and theoretical precision of each contribution.
        A 10\% uncertainty is assigned to single \PQt quark production, while a 20\% uncertainty is assigned to the remaining simulated non-\ttbar backgrounds, including \Vjets, \ttV, and \VV production.
        The QCD multijet background is estimated from data and assigned a 100\% normalization uncertainty, independently for each lepton channel and data-taking year.
        The statistical uncertainty in the transfer factors used for the data-driven QCD estimate is included as a shape uncertainty.

  \item Parton distribution functions: The NNPDF3.1 NNLO set is used to evaluate the PDF uncertainty.
        Independent \Dcb templates are constructed for the 100 PDF replicas~\cite{Ball:2017nwa}.
        The bin-by-bin envelope of these templates with respect to the nominal prediction is used to define the PDF shape variation.
        The \alpS uncertainty in the PDF determination is evaluated using the corresponding dedicated PDF weights.

\end{itemize}

In the highest-\Dcb bins, which drive the sensitivity of the measurement, single \PQt quark production and the remaining non-\ttbar backgrounds, including \Vjets, have small yields.
The impact of their modeling uncertainties on the measurement is therefore limited.

\subsection{Experimental uncertainties}
\begin{itemize}
  \item Jet flavor tagging efficiencies: The full distribution of the \textsc{DeepJet} \PQb tagging discriminant in simulation is corrected using data-to-simulation scale factors derived separately for each data-taking era.
        These corrections account for both the tagging efficiency and the discriminant shape.
        The uncertainties associated with this calibration follow the standard CMS flavor-tagging prescription~\cite{BTV-16-002} and are decomposed into separate sources for light, \PQc, and \PQb jets.
        For \PQb and light jets, the scale factors are obtained through an iterative fit to \PQb-enriched dileptonic \ttbar and light-jet-enriched $\PZ{+}\text{jets}$ events, respectively~\cite{CMS-DP-2023-005}.
        The calibration is validated in multijet and fully hadronic \ttbar events.
        For \PQc jets, the nominal scale factor is set to unity, and the dedicated uncertainties are assigned to conservatively cover possible mismodeling of the discriminant response~\cite{BTV-16-002}.
        To allow the fit more flexibility in constraining the \PQc jet response as a function of jet \pt, these uncertainties are decorrelated across three jet \pt ranges: $20 < \pt < 50\GeV$, $50 < \pt < 100\GeV$, and $\pt > 100\GeV$.
        All variations are propagated to the final templates and treated as shape uncertainties.

  \item Jet energy scale and resolution: The uncertainty associated with the JES is evaluated by shifting the \pt of the simulated jets by their respective uncertainties~\cite{CMS:2016lmd,CMS-DP-2021-033}.
        The JES uncertainty is decomposed into several mutually uncorrelated sources, such as the absolute energy scale, relative balancing, and flavor response.
        These sources are grouped based on their correlation structure across the data-taking years.
        The simulated jet momenta are smeared to match the JER observed in data.
        The uncertainty associated with the JER is determined by varying the smearing factors within their measured uncertainties.
        The variations in jet \pt and mass resulting from the JES and JER uncertainties are propagated to \ptvecmiss to maintain the momentum balance of the event.
        Additionally, uncertainties arising from unclustered energy deposits in the \ptvecmiss calculation are considered separately.
        For each of these sources, the variations are propagated through the full analysis chain, including a reevaluation of the event selection criteria.
        Since this procedure accounts for changes in both the event acceptance and the kinematic distributions, these sources are treated as shape uncertainties.

  \item Integrated luminosity and pileup: The uncertainty in the integrated luminosity is determined solely by the calibration of the luminosity monitoring systems.
        The total integrated luminosity uncertainty is decomposed into components that are correlated and uncorrelated across the three data-taking years, following the standard prescriptions for CMS data collected in 2016--2018~\cite{Sirunyan2021,CMS:LUM-20-001}.
        These uncertainties are applied as rate variations.

        In addition to the luminosity, the uncertainty associated with the modeling of pileup interactions is evaluated by varying the total inelastic \pp cross section by $\pm$4.6\% relative to its nominal value~\cite{Sirunyan2018}.
        The resulting change in the pileup reweighting modifies the distributions of the selected events and is treated as a shape uncertainty.

  \item Lepton selection and trigger: The efficiencies associated with lepton reconstruction, identification, isolation, and trigger performance are measured in data using $\PZ\to\Pl\Pl$ events via the tag-and-probe method~\cite{Khachatryan2011}.
        To correct for differences between data and simulation, scale factors are applied to the simulated events.
        The systematic uncertainties associated with these scale factors are estimated by varying the background models, the signal mass window, and the tag selection criteria used in the measurement.
        These variations are propagated to the final results as shape uncertainties.

  \item Level-1 trigger inefficiency (prefiring): A correction is applied to account for the Level-1 trigger inefficiency due to timing shifts in the ECAL and the muon systems.
        The associated uncertainty is evaluated by varying the mistiming probability by 20\% combined with its statistical uncertainty.
\end{itemize}

\section{Results}\label{sec:results}
\subsection{Statistical model}
The branching fraction of the \Wcb decay is extracted using a binned maximum likelihood fit of the \Dcb discriminant described in Section~\ref{sec:template_score} to the data.
The likelihood function, $\mathcal{L}$, is defined as the product of the Poisson probabilities for all bins across all channels, multiplied by the probability density functions constraining the nuisance parameters:
\begin{equation} \label{eq:def_likelihood}
  \mathcal{L}(r, \theta \mid \text{data}) = \prod_{i=1}^{N_{\text{cat}}} \prod_{j=1}^{N_{\text{bins}}} \frac{\left( S_{ij}(r, \theta) + B_{ij}(r, \theta) \right)^{n_{ij}}}{n_{ij}!} \re^{-\left( S_{ij}(r, \theta) + B_{ij}(r, \theta) \right)} \prod_{k=1}^{N_{\text{np}}} \pi_k(\theta_k),
\end{equation}
where the index $i$ runs over the analysis categories ($N_{\text{cat}}$) (\eg, the SL region and DL CR, lepton channels, and data-taking eras), and $j$ runs over the bins of the discriminant distributions.
The observed number of events in the $j$-th bin of the $i$-th category is denoted by $n_{ij}$.
The expected signal and background yields, $S_{ij}$ and $B_{ij}$, are functions of the parameter of interest (POI), denoted as $r$, and the nuisance parameters $\theta$, with $N_{\text{np}}$ denoting the total number of nuisance parameters.
The term $\pi_k(\theta_k)$ represents the prior probability density function for the $k$-th nuisance parameter.
The post-fit uncertainty is determined by the combination of the prior uncertainties and the information from all fitted categories.
Depending on their kinematic dependence and correlations, nuisance parameters that contribute substantially to the pre-fit uncertainty can be constrained by the simultaneous fit, leading to smaller post-fit uncertainties.

The DL CR is enriched in \tthf events and provides direct sensitivity to the rates of the corresponding background components, allowing them to be profiled using data in the simultaneous fit.
A control region is most effective when it enhances the background processes of interest while preserving the relevant theoretical and experimental properties shared with the region used to extract the signal.
In this analysis, the \tthf processes in the DL CR and SL region arise from the same \ttbar production mechanism and therefore share most systematic uncertainties related to simulated event generation and experimental corrections.
Through the common nuisance parameters, the constraints obtained from the DL CR are propagated to the \tthf prediction in the SL region.

In the statistical model, the POI is defined as $r = \VcbAbs^2 / \VcbPDG^2$, where $\VcbPDG=0.0407\pm0.0013$ is the PDG combination of inclusive and exclusive determinations~\cite{ParticleDataGroup:2026}.
This value only defines the normalization of $r$ and does not act as an external constraint in the likelihood fit; its uncertainty is not included in the quoted measurement uncertainty.
The signal and background models assume that \PQt quarks decay exclusively to a \PW boson and a \PQb quark, corresponding to $\mathcal{B}(\PQt\to\PW\PQb)=1$.
Assuming the SM couplings, the signal process rate scales linearly with $r$.
To constrain the relative contributions of the signal and background processes, CKM unitarity constraints are applied.
Specifically, we enforce the unitarity of the CKM matrix rows: $\abs{V_{\PQu\PQd}}^2 + \abs{V_{\PQu\PQs}}^2 = 1$ and $\abs{V_{\PQc\PQd}}^2 + \abs{V_{\PQc\PQs}}^2 + \abs{V_{\PQc\PQb}}^2 = 1$.
In this measurement, the contribution from $\abs{V_{\PQu\PQb}}^2$ is neglected as it is significantly smaller than the experimental sensitivity.

Under the unitarity constraint, an increase in the \Wcb branching fraction implies a corresponding decrease in the branching fractions of other hadronic \PW boson decays.
The \ttbar background processes are therefore rescaled dynamically in the fit.
The background processes involving \Wcs and \Wcd decays are scaled by a factor of:
  \begin{equation}
    \text{scale} = \frac{1 - r (\VcbPDG)^2}{1 - (\VcbPDG)^2}.
  \end{equation}
In contrast, the \Wud and \Wus decays are unaffected by variations in $\abs{V_{\PQc\PQb}}^2$ under the approximation of neglecting $\abs{V_{\PQu\PQb}}^2$, and thus are not scaled.
This procedure ensures that the sum of the branching fractions remains consistent with unity while simultaneously conserving the total inclusive \ttbar cross section in the lepton$+$jets channel.
However, the resulting post-fit scale factor for the \Wcs and \Wcd contributions is very close to unity, and its effect is therefore negligible.

It should be noted that this scaling is applied only to the \ttbar process.
Other processes capable of producing \PW bosons, such as single \PQt quark production, can in principle contribute to the \Wcb signal.
However, the yields of these processes after the baseline selection are small compared to the \ttbar yield.
In the SL region, single \PQt quark production accounts for only about 4\% of the total expected yield.
Furthermore, the reconstruction algorithm, which is optimized for the topology of \ttbar events, has a low efficiency for correctly identifying hadronic \PW decays in non-\ttbar events.
Therefore, the impact of these minor processes on the extracted signal strength is estimated to be negligible.

The statistical analysis, including the template fitting and the implementation of the constraints described above, is performed using the CMS \textsc{Combine} tool~\cite{Hayrapetyan2024}.

\subsection{Fit results}
\begin{figure}[t]
  \centering
  \includegraphics[width=0.6\textwidth]{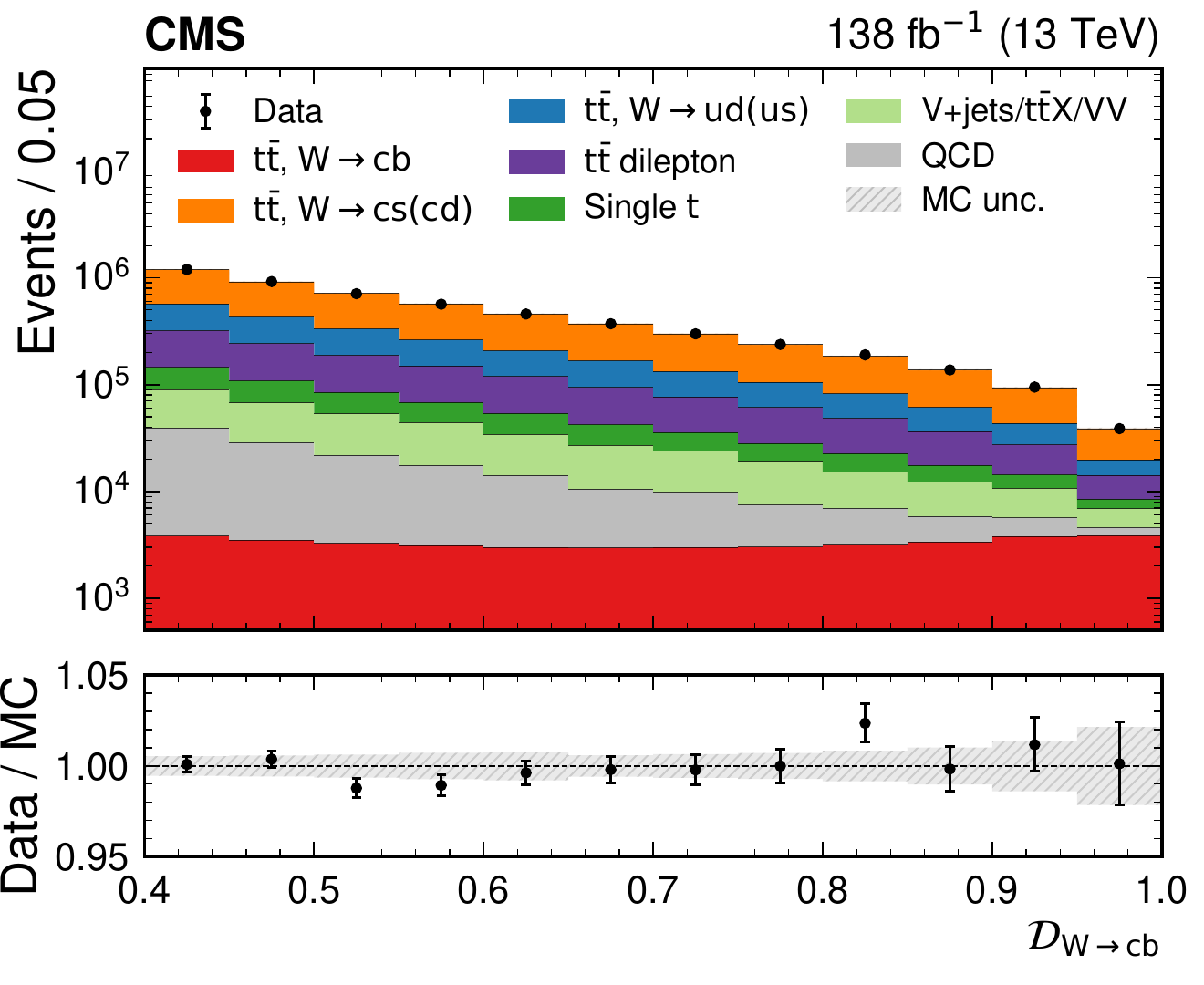}
  \caption{
    Post-fit distributions of the \Dcb discriminant in the SL region.
    For visual clarity, \tthf contributions are included in the \ttbar categories labeled by the hadronic \PW boson decay mode and in the \ttbar dilepton category.
    The vertical error bars on the data points indicate their statistical uncertainties.
    The uncertainty bands include both statistical and systematic components.
  }\label{fig:postfit_template}
\end{figure}

\begin{figure}[bt]
  \centering
  \includegraphics[width=0.6\textwidth]{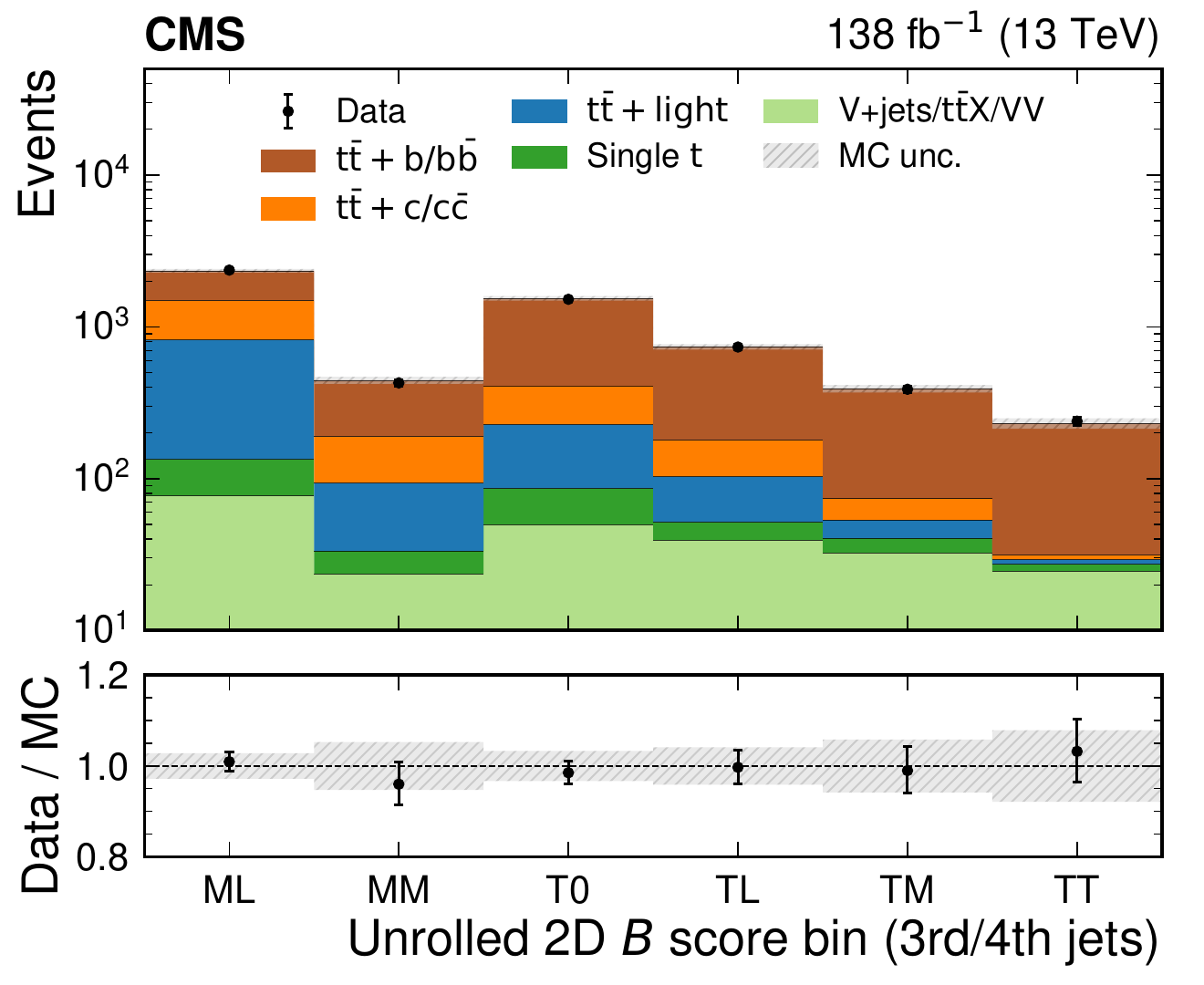}
  \caption{
    Post-fit distributions of the unrolled 2D \bvsc template in the DL CR.
    The \ttbar contributions are grouped by the flavor content of additional jets; the detailed generator-level categories are defined in Section~\ref{sec:samples}.
    The vertical error bars on the data points indicate their statistical uncertainties.
    The uncertainty bands include both statistical and systematic components.
  }\label{fig:postfit_dl}
\end{figure}

The nominal fit is performed simultaneously in the SL region and the DL CR.
The fit uses the dedicated NLO \ttbb prediction for the \ttbar background with additional \PQb-flavored jets, as described in Section~\ref{sec:samples}.
The post-fit distributions of the \Dcb discriminant in the SL region and of the unrolled 2D \bvsc template in the DL CR are shown in Figs.~\ref{fig:postfit_template} and~\ref{fig:postfit_dl}, respectively.
The post-fit distributions provide a visual check that the fitted model describes the discriminating observables.
The agreement is further quantified with a saturated goodness-of-fit test~\cite{Baker:1984} using pseudo data sets generated under the signal-plus-background hypothesis.
The test gives a $p$-value of 0.626, indicating no significant disagreement between the fitted model and the observed data.

\begin{figure}[hbt]
  \centering
  \includegraphics[width=0.6\textwidth]{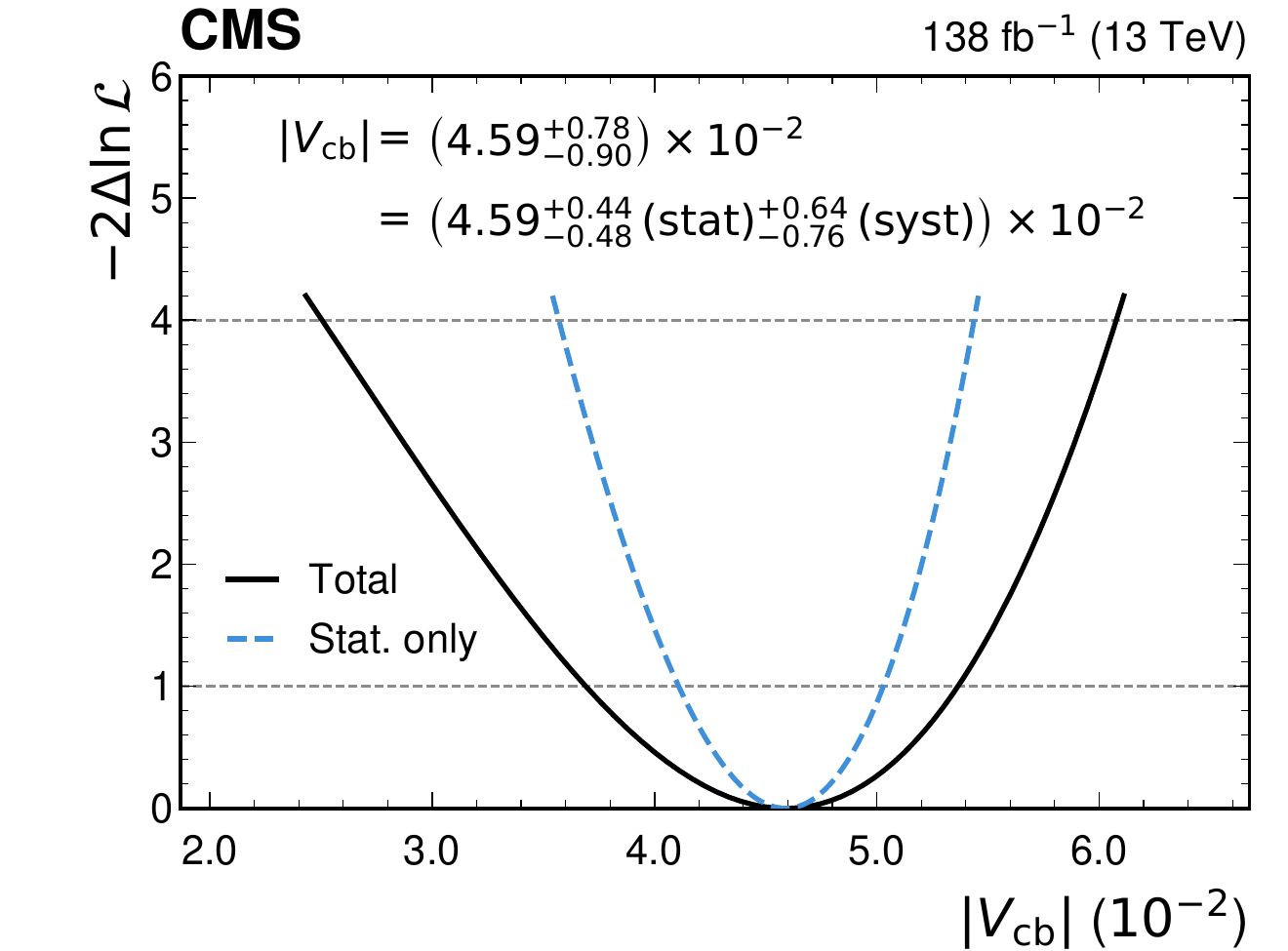}
  \caption{
    Profile likelihood scan as a function of \VcbAbs.
    The horizontal lines at $-2\Delta\ln\mathcal{L}=1$ and~4 indicate the 68\% and 95\% confidence intervals, respectively.
  }\label{fig:likelihood-scan}
\end{figure}

The measured (expected) signal strength is
  \begin{equation}
    r = 1.27^{+0.46}_{-0.44}\,\bigl(1.00^{+0.42}_{-0.41}\bigr).
  \end{equation}
The measured signal strength corresponds to
  \begin{equation}
    \VcbAbs = \bigl[4.59^{+0.44}_{-0.48}\stat^{+0.64}_{-0.76}\syst\bigr] \times 10^{-2}.
  \end{equation}
The corresponding expected result, obtained from an Asimov data set~\cite{Cowan:2010js}, is
  \begin{equation}
    \VcbAbs = \bigl[4.07^{+0.37}_{-0.40}\stat^{+0.70}_{-0.86}\syst\bigr] \times 10^{-2}.
  \end{equation}
The profile likelihood scan as a function of \VcbAbs is shown in Fig.~\ref{fig:likelihood-scan}.
The post-fit normalization factor of the combined additional-\PQb component, which is allowed to float freely in the fit, is $1.01^{+0.09}_{-0.07}$.

The leading contributions to the uncertainty in $r$ are estimated using a post-fit covariance-based uncertainty decomposition.
The five largest contributions from nuisance parameter groups are summarized in Table~\ref{tab:leading_uncertainties}.
The statistical contribution is not included in the ranking.
The precision of the measurement is primarily limited by the theoretical modeling of \ttbar production with additional HF jets.

\begin{table}[htbp]
  \centering
  \topcaption{
    Leading contributions to the uncertainty in the fitted signal strength $r$.
    Each contribution is evaluated from profile likelihood scans by fixing the corresponding group of nuisance parameters.
    Because correlations between different nuisance parameter groups cannot be uniquely assigned in this procedure, the individual contributions are not expected to add in quadrature to the total uncertainty.
  }\label{tab:leading_uncertainties}
  \begin{tabular}{lc}
    Source                                 & Uncertainty in $r$ \\
    \hline
    \muR and \muF                          & $+$0.25/$-$0.24    \\
    Additional-\PQb normalization/modeling & $+$0.15/$-$0.16    \\
    \PQb tagging                           & $+$0.12/$-$0.11    \\
    JES/JER                                & $+$0.11/$-$0.08    \\
    PS modeling                            & $+$0.10/$-$0.06    \\[\cmsTabSkip]
    Total    (statistical and systematic)  & $+$0.46/$-$0.44    \\
  \end{tabular}
\end{table}

\subsection{Comparison of results obtained with the 4FS and 5FS descriptions}
The nominal result uses the dedicated NLO \ttbb prediction in the 4FS, in which the additional \PQb quarks are included at the ME level.
To assess the dependence of the extraction on this modeling choice, the fit is repeated using the inclusive \ttbar prediction generated at NLO in QCD in the 5FS, in which additional \PQb-flavored jets are described by the PS.
The measured (expected) signal strength obtained using the 5FS description is
  \begin{equation}
    r_{\mathrm{5FS}} = 1.08^{+0.42}_{-0.41}\,\bigl(1.00^{+0.34}_{-0.33}\bigr).
  \end{equation}

Additional checks are performed to quantify whether the choice of \ttbar modeling description induces a significant bias or a preference in the data.
The cross-model Asimov fits, in which pseudo data sets generated with one prediction are fitted with the other, give shifts in $r$ that are small compared with the fit uncertainty.
Conditional log-likelihood-ratio ensembles are used to compare the two non-nested likelihood descriptions, and saturated goodness-of-fit tests are performed for the two model choices.
These tests do not indicate a significant preference of the observed data for either description.

\begin{figure}[htbp]
  \centering
  \includegraphics[width=0.9\textwidth]{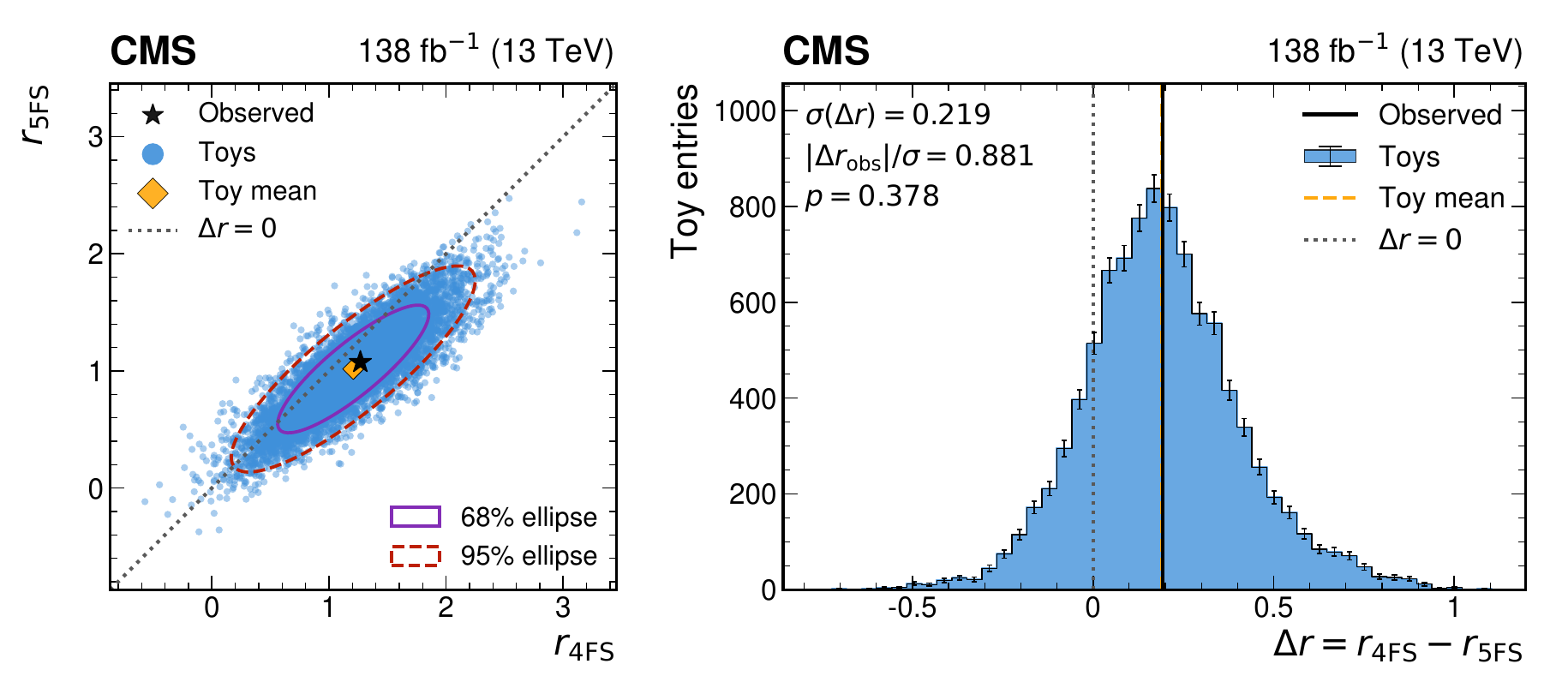}
  \caption{
    Bootstrap compatibility test of the POI values obtained with the 4FS and 5FS fits.
    The left panel shows the paired $(r_{\mathrm{4FS}}, r_{\mathrm{5FS}})$ distribution, and the right panel shows the distribution of $\Delta r = r_{\mathrm{4FS}} - r_{\mathrm{5FS}}$.
  }\label{fig:bootstrap-compatibility-4fs-5fs}
\end{figure}

Since the two fits are performed on the same data set, the fitted POI values are highly correlated.
Their compatibility is therefore further evaluated with a bootstrap test~\cite{Efron:1979}, shown in Fig.~\ref{fig:bootstrap-compatibility-4fs-5fs}.
Pseudo data sets are generated by Poisson-fluctuating the observed bin counts and are fitted with both model choices.
The resulting distribution of $\Delta r = r_{\mathrm{4FS}} - r_{\mathrm{5FS}}$ is used to account for the correlation between the two fits.
The observed difference is $\Delta r_{\text{obs}}=0.193$, while the bootstrap distribution gives $\sigma(\Delta r)=0.219$.
This corresponds to $\abs{\Delta r_{\text{obs}}}/\sigma(\Delta r)=0.881$, giving a two-sided $p$-value of 0.378.
The difference between the two results is therefore not interpreted as a significant tension.

\section{Summary}\label{sec:summary}
A measurement of \VcbAbs, the magnitude of the Cabibbo--Kobayashi--Maskawa matrix element \Vcb, has been presented using top quark pair (\ttbar) events in the lepton$+$jets final state produced in proton-proton collisions at $\sqrt{s} = 13\TeV$.
The data correspond to an integrated luminosity of 138\fbinv collected by the CMS experiment.
The analysis exploits the hadronic decay of the \PW boson in \ttbar events to determine \VcbAbs at the electroweak scale, without relying on the nonperturbative quantum chromodynamics inputs required in semileptonic \PB meson decay measurements.
The \ttbar system is reconstructed using a boosted decision tree algorithm, and a dedicated discriminant is constructed to separate the \Wcb signal from the dominant \ttbar backgrounds using flavor-tagging and kinematic information.
A simultaneous binned maximum likelihood fit is applied to the single-lepton region and the dilepton control region to extract the signal strength.
Backgrounds from \ttbar production with additional heavy-flavor (HF) jets are constrained in the fit, with the nominal modeling of the additional-\PQb component based on a dedicated next-to-leading-order \ttbb prediction in the four-flavor scheme.
An alternative fit using the inclusive \ttbar prediction at next-to-leading order in the five-flavor scheme is found to be compatible with the nominal result within the sensitivity of the analysis.

The measured signal strength is $r = 1.27^{+0.46}_{-0.44}$, corresponding to
\begin{equation}
  \VcbAbs = \bigl[4.59^{+0.44}_{-0.48}\stat^{+0.64}_{-0.76}\syst\bigr] \times 10^{-2}.
\end{equation}
The uncertainties include both statistical and systematic contributions and are dominated by the modeling of \ttbar production with additional HF jets and by flavor-tagging uncertainties.
The measured value is consistent within uncertainties with both the inclusive and exclusive determinations from semileptonic \PB meson decays~\cite{ParticleDataGroup:2026}.
Although its precision is currently lower than that of those determinations, it provides an independent determination of \VcbAbs through the decay of an on-shell \PW boson at the electroweak scale.
The strategy established here could help future studies probe the long-standing \VcbAbs discrepancy.

\begin{acknowledgments}
We congratulate our colleagues in the CERN accelerator departments for the excellent performance of the LHC and thank the technical and administrative staffs at CERN and at other CMS institutes for their contributions to the success of the CMS effort. In addition, we gratefully acknowledge the computing centers and personnel of the Worldwide LHC Computing Grid and other centers for delivering so effectively the computing infrastructure essential to our analyses. Finally, we acknowledge the enduring support for the construction and operation of the LHC, the CMS detector, and the supporting computing infrastructure provided by the following funding agencies: SC (Armenia), BMFWF and FWF (Austria); FNRS and FWO (Belgium); CNPq, CAPES, FAPERJ, FAPERGS, and FAPESP (Brazil); MES and BNSF (Bulgaria); CERN; CAS, MoST, and NSFC (China); MINCIENCIAS (Colombia); MSES and CSF (Croatia); RIF (Cyprus); SENESCYT (Ecuador); ERC PRG and PSG, TARISTU24-TK10 and MoER TK202 (Estonia); Academy of Finland, MEC, and HIP (Finland); CEA and CNRS/IN2P3 (France); SRNSF (Georgia); BMFTR, DFG, and HGF (Germany); GSRI (Greece); MATE and NKFIH (Hungary); DAE and DST (India); IPM (Iran); SFI (Ireland); INFN (Italy); MSIT and NRF (Republic of Korea); MES (Latvia); LMTLT (Lithuania); MOE and UM (Malaysia); BUAP, CINVESTAV, CONACYT, LNS, SEP, and UASLP-FAI (Mexico); MOS (Montenegro); MBIE (New Zealand); PAEC (Pakistan); MSHE, NSC, and NAWA (Poland); FCT (Portugal); MESTD (Serbia); MICIU/AEI and PCTI (Spain); MOSTR (Sri Lanka); Swiss Funding Agencies (Switzerland); NSTC (Taiwan); MHESI (Thailand); TUBITAK and TENMAK (T\"{u}rkiye); NASU (Ukraine); STFC (United Kingdom); DOE and NSF (USA).

\begin{sloppypar}
\setlength\emergencystretch{\hsize}
\hyphenation{Rachada-pisek} Individuals have received support from the Marie-Curie program and the European Research Council and Horizon 2020 Grant, contract Nos.\ 675440, 724704, 752730, 758316, 765710, 824093, 101115353, 101002207, 101001205, and COST Action CA16108 (European Union); the Leventis Foundation; the Alfred P.\ Sloan Foundation; the Alexander von Humboldt Foundation; the Science Committee, project no. 22rl-037 (Armenia); the Fonds pour la Formation \`a la Recherche dans l'Industrie et dans l'Agriculture (FRIA) and Fonds voor Wetenschappelijk Onderzoek contract No. 1228724N (Belgium); the Beijing Municipal Science \& Technology Commission, No. Z191100007219010, the Fundamental Research Funds for the Central Universities, the Ministry of Science and Technology of China under Grant No. 2023YFA1605804, the Natural Science Foundation of China under Grant No. 12535004, and USTC Research Funds of the Double First-Class Initiative No.\ YD2030002017 (China); the Ministry of Education, Youth and Sports (MEYS) of the Czech Republic; the Shota Rustaveli National Science Foundation (Georgia); the Deutsche Forschungsgemeinschaft (DFG), among others, under Germany's Excellence Strategy -- EXC 2121 ``Quantum Universe" -- 390833306, and under project number 400140256 - GRK2497; the Hellenic Foundation for Research and Innovation (HFRI), Project Number 2288 (Greece); the Hungarian Academy of Sciences, the New National Excellence Program - \'UNKP, the NKFIH research grants K 131991, K 138136, K 143460, K 143477, K 147557, K 146913, K 146914, K 147048, TKP2021-NKTA-64, 2026-4.1.1-MISSZI\'{O}K-2026-00005 and 2025-1.1.5-NEMZ\_KI-2025-00004, and MATE KKP and KKPCs Research Excellence and Flagship Research Groups grants (Hungary); the Council of Science and Industrial Research, India; ICSC -- National Research Center for High Performance Computing, Big Data and Quantum Computing, FAIR -- Future Artificial Intelligence Research, and CUP I53D23001070006 (Mission 4 Component 1), funded by the NextGenerationEU program, the Italian Ministry of University and Research (MUR) under Bando PRIN 2022 -- CUP I53C24002390006, PRIN PRIMULA 2022RBYK7T (Italy); the Latvian Council of Science; the Ministry of Science and Higher Education, project no. 2022/WK/14, and the National Science Center, contracts Opus 2021/41/B/ST2/01369, 2021/43/B/ST2/01552, 2023/49/B/ST2/03273, and the NAWA contract BPN/PPO/2021/1/00011 (Poland); the Funda\c{c}\~ao para a Ci\^encia e a Tecnologia (Portugal); the National Priorities Research Program by Qatar National Research Fund; MICIU/AEI/10.13039/501100011033, ERDF/EU, ``European Union NextGenerationEU/PRTR", projects PID2022-142604OB-C21, PID2022-139519OB-C21, PID2023-147706NB-I00, PID2023-148896NB-I00, PID2023-146983NB-I00, PID2023-147115NB-I00, PID2023-148418NB-C41, PID2023-148418NB-C42, PID2023-148418NB-C43, PID2023-148418NB-C44, PID2024-158190NB-C22, RYC2021-033305-I, RYC2024-048719-I, CNS2023-144781, CNS2024-154769 and Plan de Ciencia, Tecnolog{\'i}a e Innovaci{\'o}n de Asturias, Spain; the Chulalongkorn Academic into Its 2nd Century Project Advancement Project, the National Science, Research and Innovation Fund program IND\_FF\_68\_369\_2300\_097, and the Program Management Unit for Human Resources \& Institutional Development, Research and Innovation, grant B39G680009 (Thailand); the Eric \& Wendy Schmidt Fund for Strategic Innovation through the CERN Next Generation Triggers project under grant agreement number SIF-2023-004; the Kavli Foundation; the Nvidia Corporation; the SuperMicro Corporation; the Welch Foundation, contract C-1845; and the Weston Havens Foundation (USA).
\end{sloppypar}
\end{acknowledgments}\section*{Data availability} Release and preservation of data used by the CMS Collaboration as the basis for publications is guided by the  \href{https://doi.org/10.7483/OPENDATA.CMS.1BNU.8V1W}{CMS data preservation, re-use and open access policy}.

\bibliography{auto_generated}

\clearpage
\appendix
\section{\texorpdfstring{Optimization of class-dependent output weights for the \Dcb discriminant}{Optimization of class-dependent output weights for the Dcb discriminant}}\label{app:class_weights}
The event classifier distinguishes the seven classes defined in Table~\ref{tab:definition_classification}.
The class composition used for training differs substantially from the expected process composition in the analysis regions.
If the training events were weighted according to the expected process yields, the much lower yields of the signal classes relative to the background classes would make their contribution to the loss function negligible, potentially leading to unstable training.
We therefore train the multiclass classifier with approximately balanced training weights.
After training, separate class-dependent output weights are introduced and optimized to maximize the expected sensitivity, accounting for both the expected yields and their uncertainties in the analysis regions.

The standard softmax function, commonly used in the final output layer of a deep learning classifier, converts its unnormalized outputs into probabilities for mutually exclusive classes.
For an event $x$, let $z_i(x)$ denote the unnormalized output for class $i$.
The standard softmax output is
  \begin{equation}
    P_i(x) = \frac{\re^{z_i(x)}}{\sum_{j=1}^{C} \re^{z_j(x)}},
  \end{equation}
where $i=1,\ldots,C$ labels the classes defined in Table~\ref{tab:definition_classification}.
To adjust the class outputs after training, positive weights $w_i$ are applied as offsets $\log w_i$ to the unnormalized outputs.
The resulting weighted softmax outputs are
  \begin{equation}
    P_i^{w}(x) =
    \frac{w_i \re^{z_i(x)}}{\sum_{j=1}^{C} w_j \re^{z_j(x)}}.
  \end{equation}
The \Dcb discriminant is then defined as the sum of the weighted softmax outputs for the two signal classes,
  \begin{equation}
    \Dcb(x) = P_1^{w}(x) + P_2^{w}(x),
  \end{equation}
where Classes 1 and 2 are the matched and unmatched \Wcb signal classes, respectively.

The weights $w_i$ are optimized using an expected sensitivity metric that accounts for both statistical uncertainties and the dominant systematic uncertainties.
This is necessary because several background processes, such as \tthf, have sizable normalization uncertainties, and some systematic variations are enhanced in \PQb-enriched event topologies.
As a result, the optimal discriminant for the measurement is not necessarily obtained by a simple correction for the nominal class priors alone.

To estimate the sensitivity, we construct an approximate covariance matrix for a binned template with $N$ bins:
  \begin{equation}
    \Sigma = D + \Delta \Delta^{\mathrm{T}}.
  \end{equation}
Here,
  \begin{equation}
    D = \mathrm{diag}\left(\mu_i + \sigma^2_{\mathrm{MC},i}\right)
  \end{equation}
contains the uncorrelated bin-by-bin variances.
The Poisson variance is approximated by the nominal expected yield $\mu_i$, and $\sigma^2_{\mathrm{MC},i}$ denotes the additional uncorrelated variance arising from the limited number of simulated events.
The matrix $\Delta$ has dimensions $N{\times}K$, where each column corresponds to the absolute bin-by-bin yield shift induced by a one-standard-deviation variation of one of the $K$ nuisance parameters with the largest expected impacts, after symmetrization where applicable.

Under this Gaussian approximation, the expected sensitivity to the signal component is estimated as
  \begin{equation}
    Z^2 = \boldsymbol{s}^{\mathrm{T}} \Sigma^{-1} \boldsymbol{s},
  \end{equation}
where $\boldsymbol{s}$ is the vector of expected signal yields in the bins.
This metric accounts for both uncorrelated statistical fluctuations and correlated shape variations from systematic uncertainties.

The weight vector $\boldsymbol{w}=(w_1,\ldots,w_C)$, where $C$ is the number of classes, defines a multidimensional optimization problem.
We use JADE, an adaptive differential-evolution algorithm~\cite{Zhang:2009jade}, to find the set of weights that maximizes the sensitivity metric $Z^2$.
Since multiplying all class weights by a common constant leaves the weighted softmax outputs unchanged, one overall scale must be fixed.
In this analysis, the weights of the two signal classes are fixed to unity, and the remaining background-class weights are optimized.

The repeated evaluation of $Z^2$ during the optimization would naively require the inversion of the full $N{\times}N$ covariance matrix.
To reduce the computational cost, we use the Woodbury matrix identity,
  \begin{equation}
    \Sigma^{-1} = D^{-1} - D^{-1}\Delta \left(I_K+\Delta^{\mathrm{T}}D^{-1}\Delta\right)^{-1} \Delta^{\mathrm{T}}D^{-1}.
  \end{equation}
Here, $I_K$ is the $K{\times}K$ identity matrix.
This reduces the matrix inversion from an $N{\times}N$ matrix in the bin space to a $K{\times}K$ matrix in the nuisance parameter space, where $K\ll N$.
\cleardoublepage \section{The CMS Collaboration \label{app:collab}}\begin{sloppypar}\hyphenpenalty=5000\widowpenalty=500\clubpenalty=5000\input{TOP-26-001-public-authorlist.tex}\end{sloppypar}
%%% END EDITABLE REGION %%%
% skeleton_end
\end{document}

%% file: TOP-26-001-public-authorlist.tex
\cmsinstitute{Yerevan Physics Institute, Yerevan, Armenia}
{\tolerance=6000
A.~Gevorgyan\cmsorcid{0000-0003-2751-9489}, A.~Hayrapetyan, V.~Makarenko\cmsorcid{0000-0002-8406-8605}, A.~Tumasyan\cmsAuthorMark{1}\cmsorcid{0009-0000-0684-6742}
\par}
\cmsinstitute{Marietta Blau Institute for Particle Physics, Vienna, Austria}
{\tolerance=6000
W.~Adam\cmsorcid{0000-0001-9099-4341}, L.~Benato\cmsorcid{0000-0001-5135-7489}, T.~Bergauer\cmsorcid{0000-0002-5786-0293}, M.~Dragicevic\cmsorcid{0000-0003-1967-6783}, S.~Gundacker\cmsorcid{0000-0003-2087-3266}, A.K.~Guven\cmsorcid{0009-0004-5670-5138}, M.~Jeitler\cmsAuthorMark{2}\cmsorcid{0000-0002-5141-9560}, N.~Krammer\cmsorcid{0000-0002-0548-0985}, A.~Li\cmsorcid{0000-0002-4547-116X}, D.~Liko\cmsorcid{0000-0002-3380-473X}, M.~Matthewman, A.~Pfeiffer\cmsorcid{0000-0001-5328-448X}, J.~Schieck\cmsAuthorMark{2}\cmsorcid{0000-0002-1058-8093}, R.~Sch\"{o}fbeck\cmsAuthorMark{2}\cmsorcid{0000-0002-2332-8784}, M.~Shooshtari\cmsorcid{0009-0004-8882-4887}, N.~Van~Den~Bossche\cmsorcid{0000-0003-2973-4991}, W.~Waltenberger\cmsorcid{0000-0002-6215-7228}, C.E.~Wulz\cmsAuthorMark{2}\cmsorcid{0000-0001-9226-5812}
% ...
\par}
$^{1}$Also at Yerevan State University, Yerevan, Armenia\\
$^{2}$Also at Technische Universit\"{a}t Wien, Vienna, Austria\\
% ...
